\documentclass[12pt]{article} 
\usepackage{amsmath}
\usepackage{amsfonts}
\usepackage{amssymb}
\usepackage{graphicx}
\usepackage{color}
\usepackage{wrapfig}
\usepackage{pifont}
\usepackage{epstopdf}
\usepackage{booktabs}

\newtheorem{definition}{Definition}[section]

\newtheorem{theorem}{Theorem}[section]

\newtheorem{proposition}{Proposition}[section]

\newcommand{\cD}{\mathcal{D}}

\newenvironment{proofproof}{{\noindent\bf Proof\ }}{\hfill{$\Box$}\vspace{0.1in}}

\title{
 Locating a Tree in a Reticulation-Visible  Network in Cubic Time }
\author{Andreas D.M. Gunawan\thanks{Department of Mathematics, National University of Singapore, Singapore 119076, Singapore.}, Bhaskar DasGupta\thanks{Department of Computer Science, University of Illinois at Chicago, Chicago, IL 60607, USA. This work was supported by NSF grant IIS-1160995.}, Louxin Zhang\thanks{To whom correspondence should be addressed. Department of Mathematics, National University of Singapore, Singapore 119076, Singapore. E-mail: matzlx@nus.edu.sg. }}
\date{}

\begin{document}
\maketitle

\begin{abstract}
	In phylogenetics, phylogenetic trees are rooted binary trees,  whereas phylogenetic networks are rooted acyclic digraphs.   Edges are directed away from the root and leaves are uniquely labeled with taxa in phylogenetic networks. For the purpose of evolutionary model validation, biologists check whether or not  a  phylogenetic tree is contained in a  phylogenetic network on the same taxa. Such a tree containment problem is NP-complete. A phylogenetic network is reticulation-visible if every reticulation node separates the network root from some leaves. We answer an open problem by proving that the problem is solvable in cubic time for reticulation-visible phylogenetic networks. The key gadget used in our answer can also allow us to design a linear-time algorithm for the cluster containment problem for networks of this type and to prove that every galled network with $n$ leaves has $2(n-1)$ reticulation nodes at most.
\end{abstract}

\section{Introduction}
	How life came to existence and evolved  has been a key scientific question in the past hundreds of years.  
Traditionally, a (phylogenetic) tree has been used to model the evolutionary history of species, in which  a node  represents a \emph{speciation event} and the leaves represent the extant species under study. Such evolutionary trees are often reconstructed from the gene or protein sequences sampled from the extant  species under study. Since genomic studies have demonstrated that genetic material is often transfered between organisms in a non-reproductive manner \cite{Chan_13_PNAS,Treangen_11_PLOSGenentics}, it has been commonly accepted that (phylogenetic) networks are more suitable for modeling horizontal gene transfer, introgression, recombination and hybridization events in genome evolution \cite{Dagan_08_PNAS,Doolittle,Gusfield_14_Book,Moret_04_TCBB,Nakhleh_13_TREE}. Mathematically, a network is a rooted acyclic digraph with labeled leaves. 
Algorithmic and combinatorial aspects of  networks have been intensively studied in the past two decades \cite{Gusfield_14_Book,Huson_book,Wang_01_JCB}.
	
	An important  issue is to check the ``consistency'' of two evolutionary models. A somewhat simpler (but nonetheless very important) version of  this issue asks whether a given network is consistent with an existing tree model or not.  This motivates researchers to study the problem of determining whether a tree is displayed by a network or not, called the tree containment problem (TCP). The cluster containment problem (CCP) is another related algorithmic problem that asks whether or not a subset of taxa is a cluster in a tree displayed by a network.  Both TCP and CCP have also been investigated in the development of network metrics \cite{Cardona_09_TCBB,Kanj_08_TCS} 

The TCP and CCP are NP-complete \cite{Kanj_08_TCS},  even on a very restricted class of  networks \cite{van_Iersel_2010_IPL}.  
van Iersel, Semple and Steel posed an open problem whether or not the TCP is solvable in polynomial time for reticulation-visible networks
\cite{Gunawan_2015, Huson_book,van_Iersel_2010_IPL}. The visibility property was originally introduced to capture an important feature of galled networks \cite{Huson_Recomb}. A  network is reticulation-visible if every reticulation node separates the network root from some leaves.  Real network models are likely reticulation-visible (see  \cite{Marcussen_SysBiol_12} for example). Although great effort has been devoted to the study of the TCP, it has been shown to be polynomial-time
solvable only for a couple of very restricted subclasses of reticulation-visible networks \cite{Philippe_2015,van_Iersel_2010_IPL}. Other studies related to the TCP include \cite{Linz_Count}.
 
 In this paper,  we make three  contributions. We give an affirmative answer to the open problem by presenting a cubic time algorithm for the TCP for binary reticulation-visible networks. Additionally, we  present a linear-time algorithm for the CCP for binary reticulation-visible networks. These two algorithms are further modified into polynomial time algorithms  for non-binary reticulation-visible networks. Our algorithms rely on an important  decomposition theorem, which is proved in Section 4.  Empowered by it, we also prove that any arbitrary galled network with $n$ leaves has 
$2(n-1)$ reticulation nodes at most. 

 The rest of the paper is organized as follows. 
 Section~\ref{sec2} introduces basic concepts and notation. Section~\ref{sec3} lists our main results (Theorems ~\ref{main-theorem} and ~\ref{main-theorem2}) and gives a brief summary of algorithmic methodologies that lead us to the results. 
 In Section~\ref{sec4}, we present a decomposition theorem (Theorem ~\ref{Decomp_Thm}) that reveals an important structural property of reticulation-visible networks, based on which the two main theorems are respectively  proved in Section~\ref{sec5} and Appendix 3 (due to page limitation). 
	
\section{Basic Concepts and Notation}
\label{sec2}

\subsection{Phylogenetic networks}

In phylogenetics, \emph{networks} are acyclic digraphs in which a unique node (the \emph{root}) has a directed path to \emph{every} other node and the nodes of indegree one and outdegree zero (called the {\it leaves}) are \emph{uniquely} labeled. Leaves represent  bio-molecular sequences, extant organisms or species under study. 
In this paper, we also assume that each non-root node in a network is of  either indegree one or  outdegree one.
 A node is called a {\it reticulation} node if its indegree is strictly greater than one and its outdegree is precisely one.  Reticulation nodes represent reticulation events occurring in evolution.  
 Non-reticulation nodes are called {\it tree} nodes, which include the root and leaves.

 For convenience in describing the algorithms and proofs, we add an \emph{open} incoming edge to the root so that its degree is also 3 (Figure~\ref{example1}).
A network is  \emph{binary}  if   leaves have degree 1 and  all other   nodes have degree 3 in the network.

Let $N$ be a network. We use the following notation:\vspace{-0.5em}
\begin{itemize}
\item $\rho(N)$ is the root of $N$.\vspace{-0.5em}
\item ${\cal L}(N)$ is the set of all leaves in $N$.\vspace{-0.5em}
\item ${\cal R}(N)$ is the set of all reticulation nodes in $N$. \vspace{-0.5em}
\item 
${\cal T}(N)$ is the set of all tree nodes of degree 3 in $N$. \vspace{-0.5em}
\item ${\cal V}(N)={\cal R}(N)\cup {\cal T}(N)$, which is the set of all nodes in $N$. \vspace{-0.5em}
\item  ${\cal E}(N)$ is the set of all edges in $N$. \vspace{-0.5em}
\item For two nodes $u, v$ in $N$:
\vspace{-0.5em}
\begin{itemize}
\item $u$ is a \emph{parent} of $v$ or alternatively $v$ is a \emph{child} of $u$ if $(u, v)$ is a directed edge in $N$, and
\vspace{-0.5em}
\item 
$u$ is an \emph{ancestor} of 
$v$ or alternatively $v$ is a \emph{descendant} of $u$ if there is a directed path from $u$ to $v$. In this case, we also say $u$ is {\it below} $v$.
\end{itemize}
\vspace{-0.8em}
\item $p(u)$ is the set of the parents of $u\in {\cal R}(N)$ 
or the unique parent of $u\in {\cal T}(N)\backslash \left\{\rho(N)\right\}$.\vspace{-0.5em}
\item $c(u)$ is the set of the children for  
$u\in {\cal T}(N)$ 
or the unique child for  $u\in {\cal R}(N)$.\vspace{-0.5em}
\item ${\cal D}_{N}(u)$ is the \emph{subnetwork} vertex-induced by $u\in {\cal V}(N)$ and all descendants of $u$.
\vspace{-0.5em}
\item 
For any $E\subseteq {\cal E}(N)$, $N-E$ is the \emph{subnetwork} of $N$ with the (same) node set ${\cal V}(N)$ and the edge set 
${\cal E}(N)\backslash E$.
\vspace{-0.5em}
\item 
For any subset $V$ of nodes of $N$, $N-V$ is the \emph{subnetwork} of $N$ with
the node set ${\cal V}(N)\backslash V$ and the edge set $\{ (x, y) \in {\cal E}(N) \;|\; x\not\in V, y\not\in V\}$. 
\end{itemize}

\subsection{The visibility property}

Let $N$ be a network and $u, v\in {\cal V}(N)$.  We say that $u$ is {\it visible} (or stable) on $v$  if every path from the root $\rho(N)$ to $v$ \emph{must} contain $u$ \cite{Huson_Recomb}(also see \cite[p.~165]{Huson_book}). 
In computer science, $u$ is called a dominator of $v$ if $u$ is visible on $v$ \cite{Leng_74}.
%
%
%

A reticulation node is {\it visible} if it is visible on some leaf.  A network is {\it reticulation-visible}   if every  reticulation node is visible.  In other words, each reticulation node separates the root from 
some leaves in a reticulation-visible network. 

\begin{figure}[!t]
\begin{center}
\includegraphics[width=0.5\textwidth]{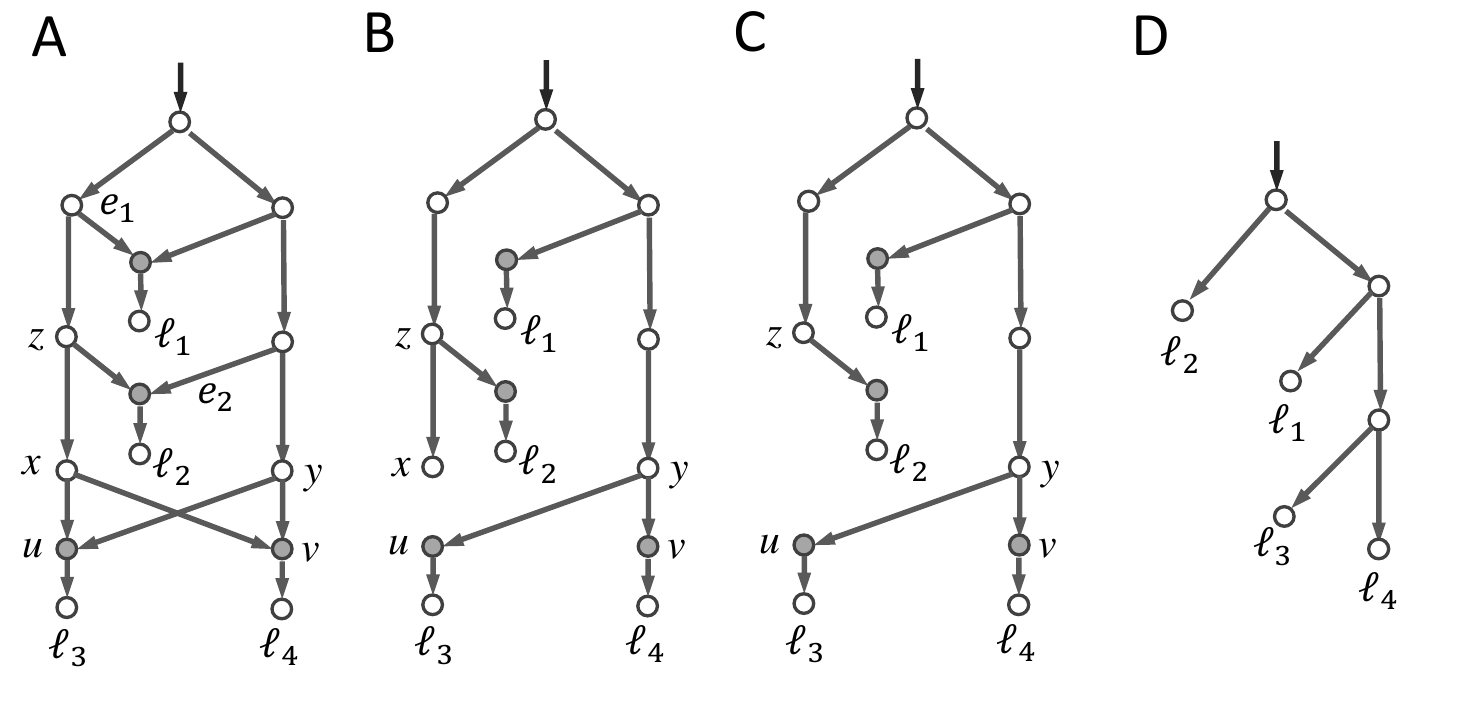}
\end{center}
\vspace*{-0.2in}
\caption{The  binary network in panel {\bf A} displays the tree in panel  {\bf D} through the removal of four edges $e_1, e_2, (x, v), (x, u)$ ({\bf B}) and the node $x$ ({\bf C}). Here, the reticulation nodes are represented by  shaded circles. \label{example1}}
\end{figure}

The phylogenetic  network in Figure~\ref{example1}A is reticulation-visible.
Clearly,  all trees are reticulation-visible, as they do not contain any reticulation nodes. In fact, 
the widely studied tree-child networks and galled networks are also reticulation-visible \cite{Cardona_09_TCBB, Wang_01_JCB}.

\subsection{The TCP and CCP}

{\em Suppression} of a node of indegree  and outdegree one  means that  the node is removed and the two edges incident to it are merged into an edge with the same orientation between the two neighbors of it. A tree $T'$  is called a {\it subdivision} of another tree $T$ if $T$ can be obtained from $T'$ by the suppression of some  nodes of indegree and outdegree one  in $T'$.

Consider a  binary network $N$ in which each reticulation node has two incoming and one outgoing edges. Thus,  the removal of one incoming edge for each reticulation node results in a directed tree.  However, there may exist new (dummy) leaves in the obtained tree. For example, after removing $e_1, e_2, (x, v)$, and $(x, u)$ in the network given in Figure~\ref{example1}A, we obtain the tree
shown in Figure~\ref{example1}B in which $x$ is a new  leaf besides the  original leaves $\ell_i$ ($1\leq i\leq 4$). If the obtained tree contains such dummy leaves,  we will have to remove them and  some of their ancestors to obtain a subtree having the \emph{same} set of leaves as $N$.

\begin{definition}[Tree Containment]
 Let $N$ be a network. For each $u\in {\cal V}(N)$, $d^{i}_u$ denotes the indegree of $u$.   $N$ {\it displays} {\rm (}or contains{\rm )} a phylogenetic  tree $T$ over the same taxa {\rm (}that is ${\cal L}(N)={\cal L}(T)${\rm )}  if   $E\subset {\cal E}(N)$ and  $V\subset {\cal V}(N)$ exist such that 
%
{\rm (i)}  $E$ contains exactly $d^{i}_u-1$ incoming edges for each 
$u\in {\cal R}(N)$,  and 
{\rm (ii)} $N-E-V$  is a subdivision of $T$.
%
\end{definition}

Because of the existence of dummy leaves,  $V$ may be nonempty to guarantee that $N-E-V$ has the same set of leaves as $T$.
 Note that a  binary network with $k$ reticulation nodes can display as many as $2^k$ trees.
%
The {\it TCP}  is to  
determine whether a  network displays a phylogenetic  tree or not.

The set of all the labeled leaves in a subtree rooted at a node  is called the \emph{cluster} of  the node  in a phylogenetic tree. 
An internal node in a network may have different clusters in different trees displayed in the network. Given a \emph{subset} of labeled leaves $B \subseteq {\cal L}(N)$, $B$ is a {\it soft cluster} in  $N$ if $B$ is the cluster of a node in some tree displayed in $N$. 
%
The {\it CCP}  is to
determine whether a  subset $B$ of ${\cal L}(N)$ is a soft cluster in a network $N$ or not. 

%
%
\section{Our Results}
\label{sec3}

\begin{theorem}[Main Result]\label{main-theorem}
Given a  binary  reticulation-visible network $N$ and a binary tree $T$, the TCP for  $N$ and $T$ can be solved in   $O(|{\cal L}(N)|^3)$ time.
\end{theorem}

The CCP is another important problem that has a quadratic-time algorithm for reticulation-visible networks \cite[pp. 168--171]{Huson_book}. 
Here, we design an optimal  algorithm for it.  The details of this algorithm  are omitted due to page limitation and 
can be found in Appendix 3. 

\begin{theorem}\label{main-theorem2}
Given a  binary reticulation-visible network $N$ and an arbitrary  subset $B$ of labeled leaves in $N$, the CCP for $N$ and $B$ can be solved in    $O(|{\cal L}(N)|)$ time. 
\end{theorem}

\noindent {\bf Synopsis of Algorithmic Methodologies}~~ 
The reader may wonder why solving the TCP is hard, as after all
$N$ is an acyclic digraph and $T$ is a binary tree. First, it is closely related to the subgraph isomorphism problem (SIP).
In general, it is very tricky to find out  whether a special case of the SIP remains NP-complete or can be solved in polynomial time. For example, whether or not  a directed tree $H$  is a subgraph of an acyclic digraph $G$ is NP-complete, but can be solved in polynomial time provided that $G$ is a forest \cite{Johnson_Book}. Second, there are non-planar reticulation-visible networks. Hence,  algorithmic techniques for planar graphs might not be suitable here. Thirdly, the TCP remains NP-complete even for  binary  networks in which  each reticulation node has a tree node as its sibling and another  tree-node as its child \cite{van_Iersel_2010_IPL}.  


 The TCP has been known to be solved in polynomial time only  for tree-child networks \cite{van_Iersel_2010_IPL} and the so called nearly-stable networks \cite{Gunawan_2015}. 
In a tree-child network,  each reticulation node is essentially connected to a leaf by a path consisting of only tree nodes. A simulation study indicates that tree-child networks comprise a very restricted  subclass of reticulation-visible networks (Gambette, http://phylnet.univ-mlv.fr/recophync/).  In a nearly-stable network,  each  child  of a node is visible if it is not visible. 
Because of the simple local structure around a reticulation node in such a network, one can determine whether or not it displays a phylogenetic tree by examining reticulation nodes one by one.
However, any approach that works on reticulation nodes one by one  is not good enough for solving the TCP for a reticulation-visible network having the structure shown in Figure~\ref{example1} if it has a larger number of reticulation nodes above the two reticulation nodes at the bottom.  We need to deal with even the whole set of reticulation nodes simultaneously for 
reticulation-visible networks of this kind.

 Our algorithms for the TCP and CCP  rely
primarily on a powerful decomposition theorem (Theorem~\ref{Decomp_Thm}). 
Roughly speaking, the theorem states that, in a reticulation-visible network,
all non-reticulation nodes can be partitioned into a collection of disjoint connected components such that 
each  component has at least {\it  two nodes} if it does not consist of a single leaf.  Most importantly,  each component  {\it contains}  either a network leaf or  all the parents of a reticulation node.

The  topological property uncovered by this theorem allows us to solve the TCP and CCP by  the divide-and-conquer
approach:  We work on the tree components one-by-one in a bottom-up fashion. In the TCP case, when working on a tree
component, we simply call a dynamic programming algorithm to decipher all the reticulation nodes right below it. In the CCP case, a slightly structural complex (but faster) dynamic programming
algorithm is used.

\section{A Decomposition Theorem}
\label{sec4}

In this section, we shall  present a decomposition theorem  that plays a vital role in designing a fast algorithm for the TCP and CCP. 
We first show that reticulation-visible networks have  two useful properties.

\begin{proposition} 
\label{basic_facts}
A reticulation-visible network $N$ has the following two properties:\vspace{-0.5em}
\begin{itemize}
\item[{\rm (a)}] {\rm (Reticulation separability)} The child and the parents of a reticulation node are tree nodes.\vspace{-0.5em}
\item[{\rm (b)}] {\rm (Visibility inheritability)} Let $E\subseteq {\cal E}(N)$.  If  $N-E$ is connected and ${\cal L}(N-E)={\cal L}(N)$,  then $N-E$ is also reticulation-visible.
\end{itemize}
\end{proposition} 
\vspace*{-0.5em}

\begin{proofproof}
(a) 
Suppose on the contrary there are $u, v\in {\cal R}(N)$ such that $v$ is the child of $u$.  
 Let $w$ be another parent of $v$.   Since $N$ is acyclic, $w$ is not below $v$ and hence not below $u$.  Since $w$ is not a descendant of $u$,   there is  a path $P(\rho(N), w)$ from $\rho(N)$ to $x$ that does not contain $u$.

We now prove that $u$ is not visible on any leaf by contradiction.
Assume $u$ is visible on a leaf $\ell$. 
 There is a path $P'$ from $\rho(N)$ to $\ell$ containing $u$. Since $v$ is the only child of $u$,  $v$ appears after $u$ in $P'$. Define $P'[v, \ell]$ to be the subpath of $P'$ from $v$ to $\ell$.  The concatenation of 
 $P(\rho(N),  w)$,  $(w, v)$, and $P'[v, \ell]$ gives a path from $\rho(N)$ to $\ell$. However, this path does not contain
$u$, a contradiction. 

\smallskip 
(b)  The proposition follows from the observation that the removal of an edge only eliminates some directed paths and does not add any new path from the root to a leaf.
\end{proofproof}


Consider a reticulation-visible network $N$. By Proposition~\ref{basic_facts}, each reticulation node is  incident to  only tree nodes. Furthermore, each connected component  $C$ of $N-{\cal R}(N)$ (ignoring edge direction) is actually a subtree of $N$ in which edges are directed away from its root.
Indeed, if $C$ contains two nodes $u$ and $v$ both of indegree 0, where indegree is defined over $N-{\cal R}(N)$, the path between $u$ and $v$ (ignoring edge direction) must contain a node $x$ with indegree $2$, contradicting that $x$ is a tree node in $N$. 
Hence, the connected components of $N- {\cal R}(N)$ are called the
{\it tree-node components} of $N$.  

\begin{figure}[!t]
\begin{center}
\includegraphics[width=0.7\textwidth]{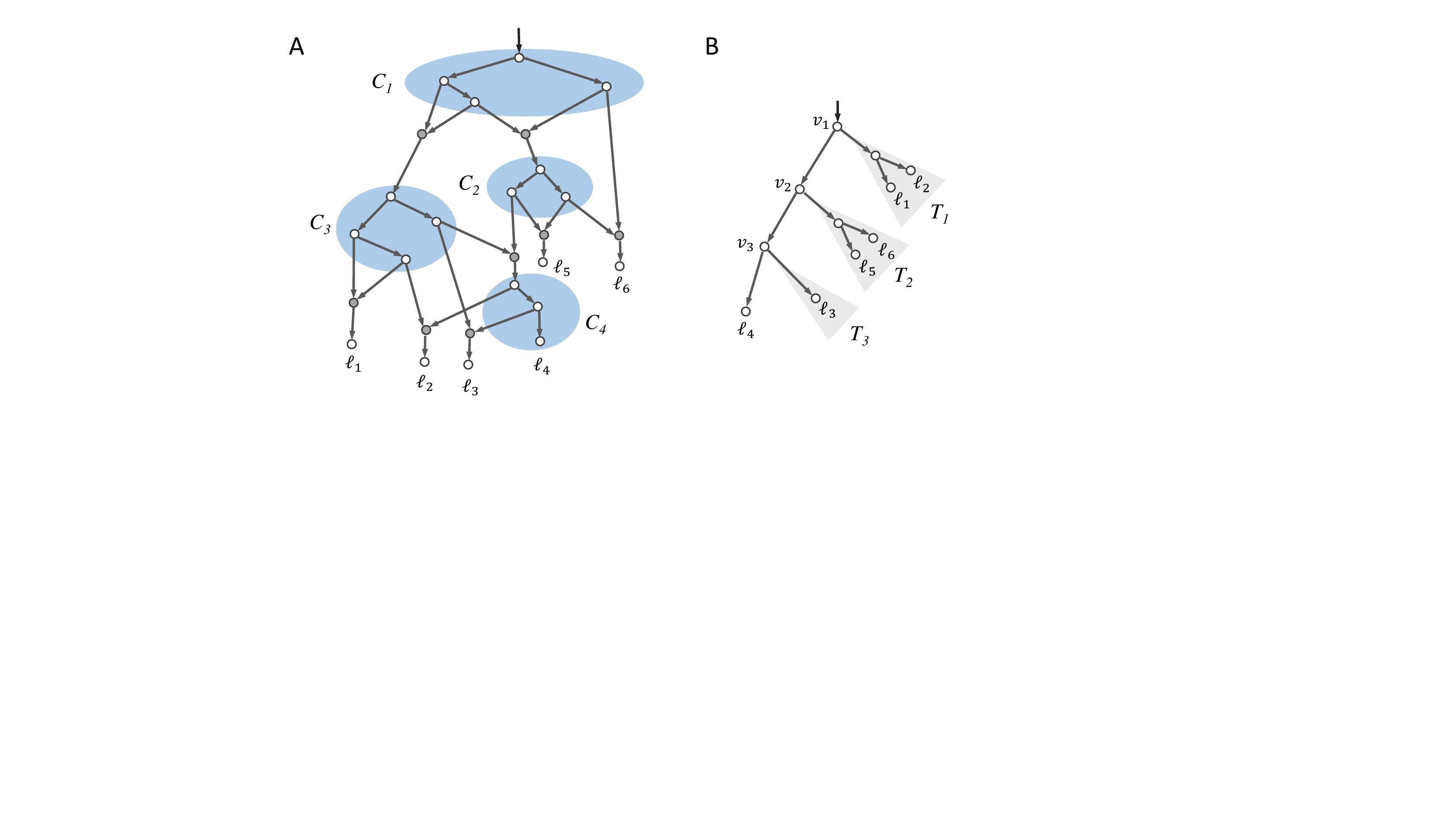}
\end{center}
\caption{({\bf A}) A binary reticulation-visible network having $9$ tree-node components, of which four ($C_1$--$C_4$) are the big ones. 
({\bf B}) A tree for consideration of its containment in the network. When  $C=C_4$ is first selected, we focus on the path from the root $v_1$ to $\ell_4$ in the given tree, where $L_{C} = \{\ell_4\}$,  $s_C = 4$ and $d_C = 3$.\label{example2}
}
\end{figure}

Let $C$ be a tree-node  component of $N$ and ${\cal V}(C)$ denote its vertex set. It
  is called a   {\it single-leaf component}  if ${\cal V}(C)=\{\ell\}$ for some $\ell\in {\cal L}(N)$.
It  is  a {\it big} tree-node component if $|{\cal V}(C)|\geq 2$.  The  binary reticulation-visible network  in Figure~\ref{example2}A has four big tree-node components and five single-leaf components.

 By definition, any two tree-node components $C'$ and $C''$ of $N$ are disjoint. We say $C'$ is \emph{below} $C''$ if there is a reticulation node $r$ such that  $C''$ contains a parent of $r$ and the child of $r$ is the root of $C'$.

\begin{theorem} 
\label{Decomp_Thm}
{\rm ({\bf Decomposition Theorem})}
Let $N$ be a reticulation-visible network with $m$ tree-node components 
$C_1, C_2, \dots, C_m$.  The following statements are true:
\begin{itemize}
\item[{\rm (i)}] ${\cal T}(N)=\uplus^{m}_{k=1}{\cal V}(C_k)$.
\vspace{-0.5em}
\item[{\rm (ii)}] For each reticulation node $r$, its child $c(r)$ is the root of some $C_i$, and each of its parents is a node in a different $C_j$. \vspace{-0.5em}

\item[{\rm (iii)}] For each tree-node component  $C_k$, \vspace{-0.5em} 
\begin{itemize} 
\item[{\rm (a)}]
$|{\cal V}(C_k)|=1$ iff  ${\cal V}(C_k)=\{\ell \}$ for some $\ell\in {\cal L}(N)$ 
{\rm (}i.e. it is  a single-leaf component{\rm )}, and \vspace{-0.5em}
\item[{\rm (b)}]
If  $C_k$ is big,  either $C_k$ contains a network leaf or   a reticulation node $r$  exists
 such that  its parents are  all in $C_k$.
 \vspace{-0.5em}
\end{itemize} 

\item[{\rm (iv)}] A big tree-node  component $C$ exists such that  there are only  single-leaf  components below it. 
\end{itemize}
\end{theorem}
\vspace*{-0.5em}
\begin{proofproof}
%
(i) The set equality  follows from the fact  that  the tree-node components are all the connected components of $N-{\cal L}(N)$.

\smallskip
(ii) Let $r\in {\cal R}(N)$. By the \emph{Reticulation Separability} property in Proposition~\ref{basic_facts}, $c(r)$ and the parents of $r$ are all  tree nodes. Thus,  by (i), each of them is in a tree-node component. 
Furthermore, since 
$c(r)$ is of indegree 0 in $N-{\cal R}(N)$, $c(r)$ must be the root of the tree-node component where it is found. 


\smallskip
(iii)
(a) Let $C$ be a tree-node component such that $|C|=1$. Assume  $C=\{u\} \subset  {\cal T}(N)\backslash {\cal L}(N)$.  
Since $u$ is the only non-leaf tree node in $C$, $p(u)\in {\cal R}(N)$ and $c(u)\subseteq {\cal R}(N)$. 
Any leaf descendant of $p(u)$ must be below some child of $u$. 
 Let $c(u)=\{c_1, c_2, \dots, c_k\}$ be the children of $u$.  Since $k$ is finite and $N$ is acyclic, there is a subset $S$
of $i$ children $c_{k_1}, c_{k_2}, \cdots,  c_{k_i}$  such that (i) $c_{k_j}$ is not below any node in $c(u)$ for each $j$, and (ii) each child
in $c(u)$ is either in $S$ or  below some child in $S$. 
For each $j\leq i$,   using the same argument as in the proof of the part (a) of  Proposition~\ref{basic_facts}, we can prove that for each leaf
$\ell$ below $c_{k_j}$, there is path from $\rho(N)$ to $\ell$ not containing $p(u)$. 
Since any leaf below $p(u)$ must be below some child in $S$,  $p(u)$ is not visible.
This contradicts the fact that $N$ is reticulation-visible. Therefore,  $|C|=1$ if and only it is a single-leaf component.

\smallskip
(b)
Assume that 
$C$ is a big tree-node component of $N$, that is, $|{\cal V}(C)| \geq 2$. 
Let $\rho(C)$ be the root of $C$.    $\rho(C)$ and its reticulation parent  are visible on some network leaf, say $\ell$.  
If $\ell$ is in $C$, we are  done. 

 If $\ell$ is not in $C $, we define ${\cal X}=\{ r\in {\cal R}(N) \;|\;  p(r)\cap C\neq \emptyset \;\;  \& \;\; \ell \mbox{ is below $r$}\}$.
For any $r', r''\in {\cal X}$, we write $r'\prec_{\cal X} r''$ if $r'$ is below $r''$, that is, there is a direct path from $r''$ to $r'$.  Since $\prec_{\cal X}$ is transitive, ${\cal X}$ is finite and  $N$ is acyclic, ${\cal X}$ contains a maximal element 
$r_m$  with respect to $\prec_{\cal X}$. Let $p(r_m)=\{p_1, p_2, \cdots, p_k\}$.  Since $r_m\in {\cal X}$, we may assume 
that $p_1\in {\cal V}(C)$. If $p_{k_0} \not\in {\cal V}(C)$ for some $1<k_0\leq k$,  $p_{k_0}$ is not below any node in $C$, as $N$ is acyclic and $r_m$ is maximal under $\prec_{\cal X}$. Hence,  there is a path $P$ from $\rho (N)$ to $p_{k_0}$ that does not contain any node in 
$C$.   Since $\ell$ is a descendant of $r_m$,  $P$ can be extended into a path from $\rho (N)$ to $\ell$ that does not contain $\rho(C)$. This contradicts that $\rho(C)$ is visible on $\ell$. Therefore, the parents of $r_m$ are all in $C$.

%

\smallskip
(iv)
It is derived from the fact that $N$ is acyclic. 
\end{proofproof}

\noindent
\textbf{Time complexity for finding tree-node decomposition} 
Let $N$ be a binary reticulation-visible  network.
Since $N$ is a DAG and has at most $8\,|{\cal L}(N)|$ nodes \cite{Gunawan_2015},  we can determine the tree-node components using the breadth-first search technique in $O(|{\cal L}(N)|)$ time.
Additionally, 
a topological ordering of its nodes can also  be found in $O(|{\cal L}(N)|)$ time. 
Using such a topological ordering, we can derive a topological ordering for the big tree-node components. With this ordering, we can identify a lowest tree-node component described in Theorem~\ref{Decomp_Thm}(iv) in constant time.

 For non-binary networks,  the above process for determining the big tree-node components and a lowest one in 
$O(|{\cal V}(N)|+|{\cal E}(N)|)$ time.

We will use the decomposition theorem to develop fast algorithms for the TCP in Section~\ref{sec5}. The theorem also seems to be very useful for studying the combinatorial aspects of networks.
A network is galled  if each reticulation node $r$ has an ancestor  $u$ such that  two internal-node disjoint paths exist  from $u$ to $r$ in which all nodes except $r$ are tree nodes.  Galled networks are reticulation-visible \cite{Huson_book}.  

\begin{theorem}
\label{bound}
 For any arbitrary galled network $N$ with $n$ leaves,    $| {\cal R} (N)| \leq 2(n-1)$.
\end{theorem}
\vspace*{-0.5em}
\begin{proofproof}
Let $N$ be a galled network with $n$ leaves. It is reticulation-visible \cite{Huson_book}. 
We consider the tree-node components of $N$. 
Since the root of each tree-node component is either the network root or the unique child of a reticulation,
\begin{eqnarray}
\label{key_eqn}
|{\cal R}(N)| =\mbox{(no. of tree-node components in $N$)} -1.
\end{eqnarray}

We first  show that $N$ does not contain any cross reticulations. 
Suppose on the contrary $N$ contains  a cross reticulation $r$. By the definition of cross reticulation, 
the parents of $r$ are in different tree-node components. 
Assume $p_1$ and $p_2$ are  two parents of $r$ in different tree-node components.  Since $N$ is acyclic, we may assume that {\it  $p_1$ is not below 
$p_2$}.   
Let $C_{p_i}$ be the tree-node components containing $p_i$ for $i=1, 2$. 
We consider the parent $r'$ of  $\rho(C_{p_2})$.  $r'$ is a reticulation node. Furthermore, $p_2$ is below $r'$ and hence $r$ is also below $r'$. 
However, we can reach $r$ from $p_1$ using a single edge without passing through $r'$, contradicting 
the Separation Lemma for galled networks \cite[p. 163]{Huson_book}.   

We have proved that $N$ does not contain any cross reticulations.  Therefore,  all the tree-node components are connected in a tree structure. Precisely, if $G$ is 
the graph whose nodes are the tree-node components  in which 
a node $X$ is connected to another $Y$ by a directed edge if the tree-node components represented by them are separated by 
a reticulation node between them, $G$ is then a rooted tree.

Consider a leaf $l$  in $G$. If the tree-node component represented by $l$ is not a single-leaf component in $N$,  there is no
reticulation node  below it and thus it must contain a leaf of $N$.   Therefore,  $G$ has at most $n$ leaves and thus at most $2n-1$ nodes. In other words, $N$ contains at most $2n-1$ tree-node components.  By Eqn.~(\ref{key_eqn}),
			$|{\cal R}(N)| \leq 2(n-1).$
\end{proofproof}

\section{A Cubic-time Algorithm for the TCP}
\label{sec5}

In this section, we first present a polynomial-time algorithm for the TCP in the case when the given reticulation-visible network is binary. Then, we describe how to modify the algorithm into one for non-binary networks. 

\subsection{An algorithm for binary networks}
 In this subsection, we assume the given network is  \emph{binary} in which a tree node has two children and a reticulation node has two parents.  We remark that each internal node has two children in a phylogenetic tree.

Let $N$ be a binary reticulation-visible network and $T$ be a  tree over the same leaves.  A reticulation node   in $N$ is {\it inner} if its parents are all in  the \emph{same} tree-node component of $N$. It is called a {\it cross} reticulation otherwise.

\begin{figure}[!t]
\begin{center}
\includegraphics[width=0.6\textwidth]{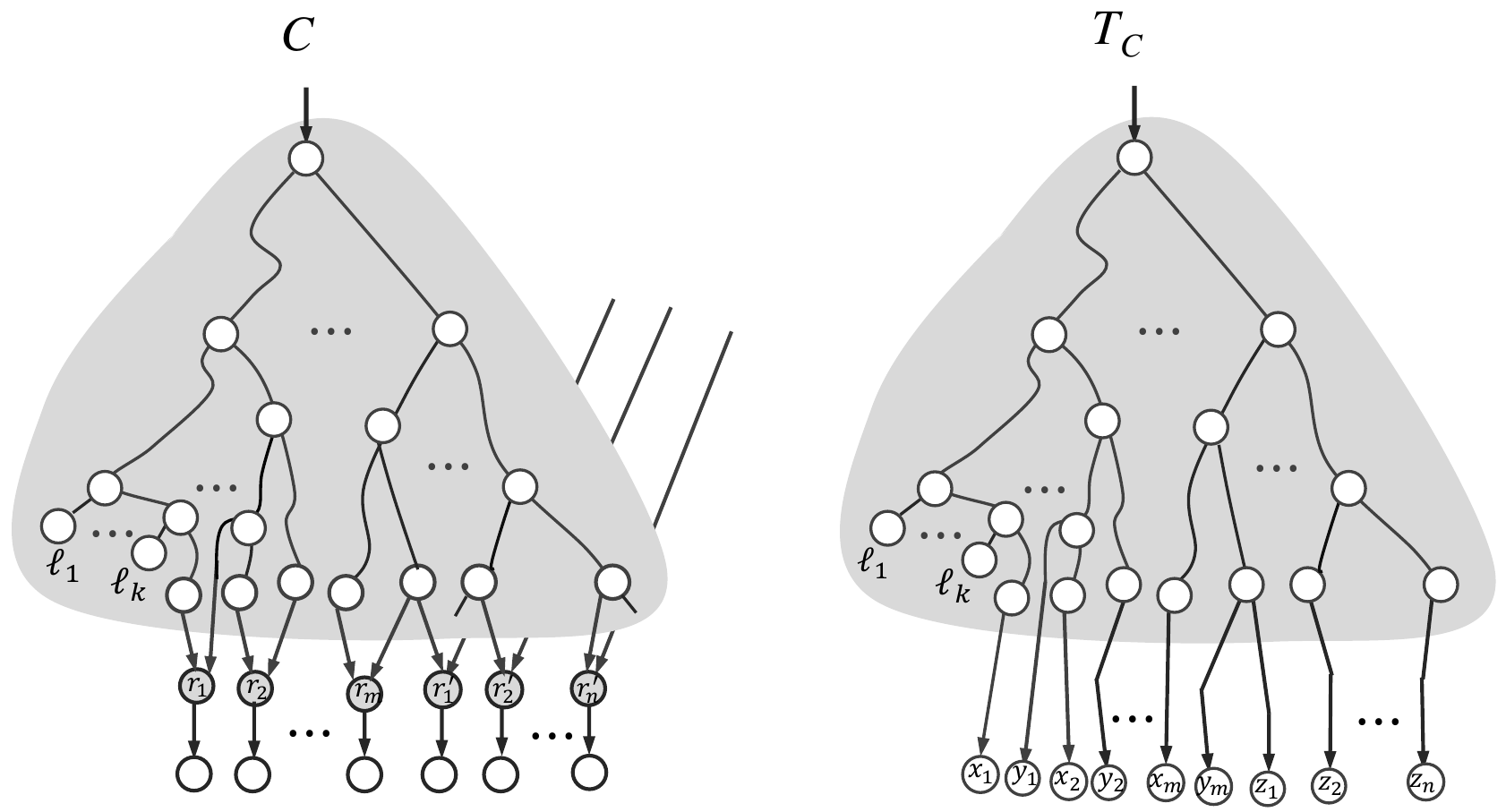}
\end{center}
\caption{Illustration of a lowest tree-node component $C$  in a binary reticulation-visible network and the corresponding tree $T_C$ 
constructed for computing $d_C$ in Proposition~\ref{Prop_55}.\label{Fig3}
}
\end{figure}

By Theorem~\ref{Decomp_Thm}, there exists a ``lowest'' big tree-node component $C$ 
below which there are only (if any) single-leaf components (Figure~\ref{Fig3}). 
We assume that $C$ contains
$k$ network leaves, say 
$\ell_1, \ell_2, \cdots, \ell_k$, and there are: \vspace{-0.5em}
\begin{itemize}
\item $m$ inner reticulations $\mbox{IR}(C)=\{r_1, r_2, \cdots, r_m\}$, and  \vspace{-0.5em}
\item $n$ cross reticulations $\mbox{CR}(C)=\{r'_1, r'_2, \cdots, r'_{m'}\}$  \vspace{-0.5em}
\end{itemize} 
below $C$.    Since  $C$ is a big tree-node components, it has two or more nodes, implying that $k+m+m'\geq 2$.   
%
%

Let $\rho(C)$ denote the root of $C$. We further define:
\begin{equation}
\label{eq0}
L_C = \left\{\,\ell_1, \ell_2, \dots, \ell_k, c(r_1), c(r_2), \cdots, c(r_m)\right\}.\end{equation}
 By Theorem~\ref{Decomp_Thm}(iii)(b), 
$k+m\geq 1$ and so $L_C$ is non-empty.   Each path $P$ from $\rho(N)$ to $c(r_i)\in L_C$ must contain $r_i$. Since the parents of $r_i$ are all in $C$, $P$ must contain $\rho(C)$. Hence,   $\rho(C)$ is visible on each network leaf in $L_C$.

We select an $\ell \in L_C$.
Since $T$ has the same leaves as $N$, $\ell \in {\cal L}(T)$ and 
there is a unique path $P_T$ from  $\rho(T)$ to $\ell$ in $T$.  Let:
\begin{equation}
\label{eq1}
P_T:\; v_1, v_2,  \dots, v_{t}, v_{t+1},
\end{equation} 
where $v_1=\rho(T)$ and $v_{t+1}=\ell$.
Then,  $T-P_T$ is a union of $t$ disjoint subtrees $ T_1, T_2,  \dots, T_{t}$, where $T_i$ is the subtree rooted at the sibling of $v_{i+1}$ for each $i=1, 2, \cdots, t$ (see Figure~\ref{example2}B). For the sake of convenience, we consider the single leaf $\ell$ as a subtree, written as $T_{t+1}$. 
We now define $s_C$ as:
 \begin{equation}
 \label{eq2}
 s_C=\min \{s  \;|\; {\cal L}(T_s)\cap L_C \neq \phi \} \end{equation} 
 Since $\ell \in {\cal L}(T_{t+1})\cap L_C$,  $s_C$ is well defined. In the example given in Figure~\ref{example2}, 
$s_C=4$.

\begin{proposition}
\label{Prop_51}
The index $s_C$ can be computed in  $O(|{\cal L}(N)|)$ time.
\end{proposition}
\vspace*{-0.5em}

\begin{proofproof}  
 Since $T$ is a binary tree with the same set of labeled leaves as the network $N$. 
$T$ has $2|{\cal L}(N)|-1$ nodes and $2|{\cal L}(N)|-2$ edges.  For each $x\in {\cal V}(T)$, we define a flag variable $f_x$ to indicate whether the subtree below $x$ contains a network leaf in $L_C$ or not.   We first traverse $T$ in the post-order: \vspace*{-0.5em}
\begin{itemize}
\item For a leaf $x\in {\cal L}(T)$, $f_x=1$ if $u\in L_C$ and 0 otherwise. \vspace*{-0.5em}
\item For an non-leaf node $x$ with children $y$ and $z$, $f_x=\max\{f_y, f_z\}$. \vspace*{-0.5em}
\end{itemize} 
Then, we compute $s_C$ as 
$s_C=\min \{ i\;|\; f_{\rho(T_i)}=1\}.$
Clearly, this algorithm correctly computes $s_C$ in $O(|{\cal L}(T)|)$ time.
\end{proofproof}

\begin{proposition}
\label{Prop52}
If $N$ displays $T$, 
then ${\cal D}_{T}\left(v_{s_C}\right)$ is displayed in ${\cal D}_N\left(\rho(C)\right)$. 
\end{proposition} 
\vspace*{-0.5em}
\begin{proofproof} 
When $s_C = t+1$, then the statement  is trivial,
as $\rho(C)$ is visible on $\ell$ and thus every path from the network root to $\ell$ must contain $\rho(C)$.

When $s_C < t+1$, by the definition of $s_C$,  there is a network leaf $\ell'$ in $ {\cal L} (T_{s_C})\cap L_C$ such that $\ell'\neq \ell$. 
If $N$ displays $T$,  $T$ has a subdivision $T'$ in $N$.  Recall that  $\rho(C)$ is visible on both $\ell$ and $\ell'$. The paths from $\rho(T')$ to $\ell$ and to $\ell'$ in $T'$ must both contain  $\rho(C)$. Since $T'$ is a tree, the lowest common ancestor $a(\ell, \ell')$ of $\ell$ and $\ell'$ is a descendant of $\rho(C)$ in $T'$ and it is the node in $T'$ that corresponds to $v_{s_C}$.  
Therefore,  the subnetwork of $T'$ below $a(\ell, \ell')$ is a subdivision of  ${\cal D}_{T}\left(v_{s_C}\right)$, that is, 
 ${\cal D}_N\left(\rho(C)\right)$ displays ${\cal D}_{T}\left(v_{s_C}\right)$.
\end{proofproof}

If $N$ displays $T$,  then $C$ may display more that ${\cal D}_{T}\left(v_{s_C}\right)$. In other words, it may display a subtree  ${\cal D}_{T}\left(v_j\right)$ for some $j < s_C$.
We define:
\begin{equation}
\label{eq3}
 d_C=\min \,\left\{ j \; |\; {\cal D}_{T}\left(v_j\right) \mbox{ is displayed in ${\cal D}_N \left(\rho(C)\right)$} \,\right\}
\end{equation}
In the example given in Figure~\ref{example2}, 
$d_C=3$.

\begin{proposition}
\label{Prop53}
If $N$ displays $T$, there must be a subdivision $T''$ of $T$ in $N$ such that 
the node {\rm (}in $T''${\rm )} corresponding to $v_{d_C}$ is in $C$. 
\end{proposition}
\vspace*{-0.5em}
\begin{proofproof} 
Assume that $N$ displays $T$ via  a subdivision $T'$ of $T$. Let  $u$ be the node in $T'$ that corresponds to $v_{d_C}$. 
Since  $\rho(C)$ is visible on the leaf $\ell$, $\rho(C)$ is in the unique path $P$ from the root to $\ell$ in $T'$. 
If $u$ is  $\rho(C)$ or below it,  we are  done. 

Assume that $u$ is neither  $\rho(C)$ nor below $\rho(C)$ in $T'$.  
 Since $\ell$ is a network leaf below $v_{d_C}$ in $T$, $\ell$ is also below $u$ in $T'$. Hence,  $P$ must also  contain $u$. 
 Since $u$ is not below $\rho(C)$,  $\rho(C)$ must be below $u$ in $P$.

On the other hand, by assumption, ${\cal D}_{T}(v_{d_C})$ is displayed in ${\cal D}_N(\rho(C))$. It has a subdivision $T^*$ in ${\cal D}_N(\rho(C))$. Let $v_{d_C}$ correspond to $u'$ in $T^*$. 
It is not hard to see that 
$u'$ is in the path from $\rho(C)$ to $\ell$ in $T^*$. 

Let $P'$ be the subpath from $u$ to 
$\rho(C)$ of $P$   
  and $P''$  be the path  from  $\rho(C)$ to $u'$ in $C$. 
Since the subtree below $u'$ in $T^*$ and the subtree below $u$ in $T'$ has the same set of labeled leaves as the subtree below $v_{d_C}$ in $T$,  
$$T'-{\cal D}_{T'}(u) + P' +P''+ 
 {\cal D}_{T^{*}}(u')$$ is also 
 a subdivision of $T$ in $N$, in which $v_{d_C}$ is mapped to $u'$ in $C$. Here, $G+H$ is the graph with the same node set as $G$ and the edge set being the union of $E(G)$ and $E(H)$ for graphs $G$ and $H$ such that $V(H) \subseteq V(G)$.
\end{proofproof}

To compute $d_C$ defined in Eqn.(\ref{eq3}),
we create a tree $T_C$ from $C$ by attaching two identical copies of the network leaf below each $r\in \mbox{IR}(C)$  to its parents in $p(r)$ in $C$ and one copy of the network leaf below 
$r\in \mbox{CR}(C)$ to the parent in $p(r)\cap {\cal V}(C)$. That is, 
 $T_C$ has  the node set:
\begin{equation}
\label{eq4}
 {\cal V}(T_C)={\cal V}(C) \cup \{ x_r, y_r \;|\;
   r\in \mbox{IR}(C)\} \cup \{ z_r \;|\; r\in \mbox{CR}(C) \}, 
\end{equation}
and the edge set
\begin{equation}
\label{eq5}
\begin{array}{lll}
 {\cal E}(T_C)&=&{\cal E}(C) \cup \{ 
 (u_r, x_r), (v_r, y_r) \;|\;
  r\in \mbox{IR}(C) \mbox{ with }  p(r)=\{u_r, v_r\}\;\}\\ && \cup \; \{ (u_r, z_r) \;|\;r\in \mbox{CR}(C)
 \mbox{ with }  \{u_r\}=p(r)\cap {\cal V}(C)\;\},
  \end{array}
\end{equation}
where $x_r$,  $y_r$, and $z_r$ are new leaves with the same label as $c(r)$ for each $r$ in $\mbox{IR}(C)$ or $ \mbox{CR}(C)$. $T_C$ is illustrated in Figure~\ref{Fig3}.

\begin{proposition}
\label{Prop_55}
There is a dynamic programming algorithm that takes $T_C$ and $T$ as input and outputs $d_C$ defined in Eqn.~(\ref{eq3}) in $O(|{\cal V}(C)|^2|{\cal V}(T)|)$ time. 
\end{proposition} 
\vspace*{-0.5em}
\begin{proofproof}
It is a special case of a problem studied in \cite{Cui_Zhang}.

For each $r\in \mbox{IR}(C)$, $T_C$ contains two leaves with the same label as $c(r)$. 
To detect whether or not $\cD_T\left(v_j\right)$ is displayed in $\cD_N\left(\rho(C)\right)$, 
we have to consider which of these two leaves will be removed. 
Such leaves will be referred to the {\it ambiguous} leaves. We use $A(T_C)$ to denote the set of ambiguous 
leaves in $T_C$.

For each $r\in \mbox{CR}(C)$, $T_C$ contains one leaf with the same label as $c(r)$. 
Similar to the case of ambiguous leaves,  this leaf may be removed or kept.  Such leaves are called 
{\it optional} leaves. We use $O(T_C)$ to denote the set of optional leaves in $T_C$.

Since each node in $C$ is a tree node of degree 3 in $N$,  $T_C$ is a full binary tree with at most $2|{\cal V}(C)|+1$ nodes. 

For our purpose, we shall present a dynamic programming algorithm to compute the following set $S_u$ of nodes in $T$:
$$ S_u =\{ x \in {\cal V}(T) \;|\;  {\cal D}_{T_C}(u) \mbox{ displays  ${\cal D}_{T}(x)$ where  $u$ is mapped to
$x$}\}$$
for each node $u$ in $T_C$.
Here, that ${\cal D}_{T_C}(u)$  displays   ${\cal D}_{T}(x)$ means that  $V_x \subseteq {\cal V}(T_C)$ exists such that 
 $T_C - V_x$ is a subdivision of ${\cal D}_{T}(x)$.
Since $T_C$ is a tree,  $V_x$ consists of  all the ambiguous or optional leaves that are in ${\cal D}_{T_C}(u)$ but not in ${\cal D}_{T}(x)$ and some ancestors of these leaves.

We introduce a boolean variable 
$f_{ux}$ to indicate whether or not ${\cal D}_{T_C}(u)$  displays   ${\cal D}_{T}(x)$  and  a set variable $M_{ux}$ to record those removed leaves if so. That is, 
\begin{eqnarray}
f_{ux}=\left\{ \begin{array}{ll}
        1 & \mbox{if ${\cal D}_{T_C}(u)$  displays   ${\cal D}_{T}(x)$,}\\
        0 & \mbox{otherwise,}
   \end{array}
 \right. \label{fux_def}
\end{eqnarray}
and 
\begin{eqnarray}
\label{mux_def}
  M_{ux}={\cal L}_{T_C}(u) \backslash  {\cal L}_{T}(x) 
\end{eqnarray}
 if $f_{ux}=1$, where ${\cal L}_Y (z)$ denotes the set of leaves below $z$ in the tree $Y$. 

When $u$ is a leaf in $T_C$,   we consider whether $x$ is a leaf or not to compute $f_{ux}$. 

If $x$ is a leaf with the same label as $u$, then ${\cal D}_{T_C}(u)$  displays   ${\cal D}_{T}(x)$ and
thus we set 
   $f_{ux}=1$  and $M_{ux}=\emptyset$.

If $x$ is a leaf and its label is different from the label of $u$,  ${\cal D}_{T_C}(u)$  does not display   ${\cal D}_{T}(x)$. In this case,   $f_{ux}=0$ and $M_{ux}$ is undefined. 

If $x$ is not a leaf, it is trivial that 
${\cal D}_{T_C}(u)$  does not display   ${\cal D}_{T}(x)$. Hence,  $f_{ux}=0$ and $M_{ux}$ is undefined.

When $u$ is an internal node with children $v$ and $w$ in $T_C$, we consider similar cases. 
If $x$ is a leaf, ${\cal D}_{T}(x)$ may or may not be displayed in ${\cal D}_{T_C}(u)$ if it is displayed below a child of $u$.  
If ${\cal D}_{T}(x)$ is displayed  below $v$,  it is also displayed at 
$u$ only if every  ambiguous leaf in $M_{vx}$ is not below $w$ and all the leaves below $w$ are either ambiguous or optional. If ${\cal D}_{T}(x)$ is displayed  below neither $v$ nor $w$, it is not displayed at $u$.  The remaining cases can be found in Table~\ref{table1}

\begin{table}[!t]
\begin{center}
{\small 
\begin{tabular}{c|c|c|l}
\toprule
  $x\in {\cal L}(T)$? & $f_{vx}$ & $f_{wx}$ & $f_{ux}$ and $M_{ux}$\\
\midrule
  Yes &$1$  &$1$ &  (a.)\\
  &&&        $ f_{ux}=\left\{\begin{array}{ll}
                    1 & \mbox{ if } M_{vx}\cap M_{wx}  = \emptyset,  \\
                    0 & \mbox{ otherwise.}
                    \end{array} \right.$  \\
           &                                   &                                    & $M_{ux}=M_{vx}\cup M_{wx} \cup \{ x \}$ if $f_{ux}=1$ \\
&&&\\
       &$1$  &$0$ &   (b.) \\
&&&  $ f_{ux}=\left\{\begin{array}{ll}
                                                                1 & \mbox{ if } {\cal L}({\cal D}_{T_C}(w)) \subseteq  A(T_C)\cup {O}(T_C),  \\
                                                                              &  \;\;\;\;  \& \; M_{vx}\cap  {\cal L}({\cal D}_{T_C}(w))   = \emptyset,\\
                                                                0 & \mbox{ otherwise.}
                                                           \end{array} \right.$   \\
 &                                   &                                    & $M_{ux}=M_{vx}\cup {\cal L}({\cal D}_{T_C}(w))$  if $f_{ux}=1$\\
&&&\\
       &$0$  &$1$ &  (c.) \\
&&&    $f_{ux}=\left\{\begin{array}{ll}
                                                                1 & \mbox{ if } {\cal L}({\cal D}_{T_C}(v)) \subseteq  A(T_C)\cup {O}(T_C),  \\
                                                                              &  \;\;\;\;  \& \; M_{wx}\cap  {\cal L}({\cal D}_{T_C}(v))     = \emptyset,\\
                                                                0 & \mbox{ otherwise.}
                                                           \end{array} \right.$   \\
 &                                   &                                    & $M_{ux}=M_{wx}\cup {\cal L}({\cal D}_{T_C}(v))$  if $f_{ux}=1$   \\
&&&\\
   &$0$  & $0$ & $f_{ux}=0$\\
\midrule
No.  &  $1$  & (d.) $1$ & $f_{ux}$ is defined same as the leaf case \\
 $c(x)=\{y, z\}$& & & $M_{ux}=M_{vx}\cup M_{wx} \cup {\cal L}(\cD_T(x))$ if $f_{ux}=1$  \\
&&&\\
     &$1$  &$0$ &  (e.)   same as the leaf case \\
&&&\\
     &$0$  &$1$ &  (f.)  same as the leaf case \\
&&&\\
&&&\\
       &$0$  &$0$ & (h.) \\
                    &&&                                $f_{ux}=\left\{\begin{array}{ll}
                                                                1 & \mbox{ if } f_{vy}=1 \;\& \; f_{wz}=1 \\
                                                  &  \;\;\;\; \& \; M_{vy} \cap  M_{wz}  =\emptyset , \\
                                                                           1 & \mbox{ if } f_{vz}=1 \;\& \; f_{wy}=1 \\
                                                  &  \;\;\;\; \& \; M_{vz} \cap  M_{wy} =\emptyset, \\       
                                                                0 & \mbox{ otherwise.}
                                                           \end{array} \right.$   \\
 &                                   &     & $M_{ux}=\left\{\begin{array}{ll} 
                                                M_{vy}\cup M_{wz} & \mbox{ if } f_{vy}=1 \;\& \; f_{wz}=1, \\
                                                M_{vz}\cup M_{wy} & \mbox{ if } f_{vz}=1 \;\& \; f_{wy}=1. \\
                                             \end{array} \right.$ \\
\bottomrule
\end{tabular} 
} 
\end{center}
\caption{The recursive formulas on $f_{ux}$ for different cases when $u$ is an internal node in $T_C$. \label{table1}}
\end{table}

Our dynamic programming algorithm  recursively computes 
$f_{ux}$ for each $u$ and $x$  by traversing both $T_C$ and $T$
in the post-order. For each $u$ and $x$, we compute $f_{ux}$ using the formulas listed in Table~\ref{table1}. Note that 
$M_{ux}$ is a subset of $A(T_C)\cup O(T_C)$ and hence has at most 
$2|{\cal V}(C)|$ elements. Therefore, each recursive step takes 
$O(|{\cal V}(C)|)$ time. 
Because of this, the total time taken by the algorithm is $O(|{\cal V}(T)|\cdot|{\cal V}(C)|^2)$.

After we know the values of $f_{ux}$ for every $u$ in $T_C$ and every $x$ in $T_c$.
we can compute $d_C$ such that ${\cal D}_{T}(v_j)$ is displayed in 
${\cal D}_{N}(\rho(C))$ as
$$ d_C= \min _{1\leq j\leq t+1} \{ j \;|\; f_{u v_j}=1 \mbox{ for some } u\in {\cal  V}(T_C) \}.$$
\end{proofproof}

\newpage 
Based on the facts presented above, we obtain the following  algorithm for the TCP. \vspace{1em} 

\begin{center}
\begin{tabular}{l}
\toprule
\hspace*{5em}{\sc The TCP Algorithm}\vspace{0.5em}\\
 Input: A reticulation-visible network $N$ and a tree $T$, which are binary.\vspace{0.5em}\\
 1. Decompose $N$ into tree-node components: $C_1\prec  C_2 \prec  \cdots \prec  C_t$, \\
 ~~~where $\prec $ is a topological order such that that  no directed path \\
~~~ from a node $u$ in $C_j$ to a node $v$ in $ C_i$ exists if $j>i$;\\
 2. $N'\leftarrow N$ and $T'\leftarrow T$;\\
 3. {\bf Repeat} unless ($N'$ becomes a single node) \{\\
 \hspace*{1em} 3.1. Select a lowest big tree-node component $C$;\\
 \hspace*{1em} 3.2. Compute $L_C$ in Eqn.~(\ref{eq0}) and select $\ell \in L_C$;\\
 \hspace*{1em} 3.3. Compute the path $P_{T}$ from the root to $\ell$  in Eqn.~(\ref{eq1});\\
\hspace*{1em} 3.4. Determine the smallest index $s_C$ defined by Eqn.~(\ref{eq2});\\
\hspace*{1em} 3.5. Determine the smallest index $d_C$ defined by Eqn.~(\ref{eq3});\\
\hspace*{1em} 3.6. {\bf If} ($s_C > d_C$), output ``$N$ does not display $T$";\\
\hspace*{1em}~~~~~~~{\bf else} \{\\
\hspace*{4em} For each $r\in \mbox{CR}(C)$ $\{$\\
\hspace*{5em} if ($c(r)\not\in \mathcal{D}_T(v_{d_C})$), delete $(z, r)$ for $z\in p(r)\cap {\cal V}(C)$;\\
\hspace*{5em} if ($c(r)\in \mathcal{D}_T(v_{d_C})$), delete $(z, r)$ for $z\in p(r)\backslash {\cal V}(C)$;\\
\hspace*{4em} $\}$\\
\hspace*{4em} Replace $C$ (resp. ${\cal D}_T(v_{d_C})$)  by a leaf $\ell_C$ in $N'$ (resp. $T'$);\\
\hspace*{4em} Remove $C$ from the list of tree-node components;\\
\hspace*{4em} Update $\mbox{CR}(C')$  for the affected big tree-node components $C'$; \\ 
\hspace*{3em} \} /* {\tt end if} */\\
\hspace*{1em} \} /* {\tt end repeat} */\\
 \bottomrule
\end{tabular}
\end{center}
\vspace*{1em}

We now analyze the time complexity of the {\sc TCP 
Algorithm}. 
Note that  $N$ has at most $8|{\cal L}(N)|$  nodes \cite{Gunawan_2015} and the input tree contains $2|{\cal L}(N)|-1$ nodes.   Step 1 can be done in $O(|{\cal L}(N)|)$ time if the breadth-first search is used. 

Step 3 is a while-loop. During each execution of this step, 
the current network is obtained from the previous network by replacing the big tree-node component examined in the last execution with a new leaf node. 
Because of this, the modification done in the last two lines in Step 3.6 makes the tree decomposition of the current network  available before the current execution. Hence, Step 3.1 takes a constant time. 
The time spent in Step 3.2 for each execution is linear in the sum of the numbers of the leaves in $C$  and of the inner reticulations below $C$. Hence, the total time spent in Step 3.2 is 
$O(|{\cal L}(N)|)$, as each reticulation is examined twice at most. 

By Proposition~\ref{Prop_51}, the total  time spent in Step 3.4  is $O(|{\cal L}(N)|^2)$. 
By Proposition~\ref{Prop_55}, the total  time spent in Step 3.5  is 
$\sum_{i} O(|{\cal V}(C_i)|^2|{\cal L}(T)|)$, which is $O(|{\cal L}(N)|^3)$. 
The time spend in Step 3.6 for each execution is $O(|{\cal V}(C)|$. Hence, the total time spent in Step 3.6 is 
$O(|{\cal L}(N)|$.
Taken altogether, the above facts imply that  the {\sc TCP Algorithm} takes  $O(|{\cal L}(N)|^3)$ time, 
proving Theorem~\ref{main-theorem}.
%

\subsection{Generalization to non-binary networks}
 
The algorithm in the last subsection can be easily modified into a polynomial time TCP algorithm for non-binary reticulation-visible networks in which tree nodes are of indegree 1 and outdegree greater than 1, whereas reticulation nodes are of  indegree greater than 1 and outdegree 1.   
First, we have proved the decomposition theorem  for arbitrary reticulation-visible networks. Second, the concepts of inner and cross reticulation remain the same. 
Third,  Propositions~\ref{Prop_51}-\ref{Prop53} still hold.  The only modification we have to make is on the definition of 
$T_C$, appearing in Step 3.5, and on the proof of Proposition~\ref{Prop_55}. It can be done as follows.

Eqn.~(\ref{eq4}) and (\ref{eq5})  now become 
\begin{eqnarray}
\label{eq44}
 {\cal V}(T_C) &=&{\cal V}(C)  \nonumber \\
   & &  \cup \; \{ x^{(i)}_r \;|\;
   r\in \mbox{IR}(C)  \;\& \; 1\leq i\leq |p(r)|\; \} \nonumber \\
   && \cup \; \{ z^{(i)}_r \;|\; r\in \mbox{CR}(C) \mbox{ having $k$ parents in $C$ } \;\& \; 1\leq i\leq k\; \}, 
\end{eqnarray}
and the edge set
\begin{eqnarray}
\label{eq55}
 {\cal E}(T_C)&=&{\cal E}(C)   \nonumber \\
    & & \cup \; \{ 
 (u^{(i)}_r, x^{(i)}_r) \;|\;
  r\in \mbox{IR}(C) \mbox{ with }  p(r)=\{u^{(i)}_r | 1\leq i\leq k \}\;\} \nonumber \\ 
&& \cup \; \{ (u^{(i)}_r, z^{(i)}_r) \;|\;r\in \mbox{CR}(C)
 \mbox{ with }  p(r)\cap {\cal V}(C)=\{u^{(i)}_r | 1\leq i\leq k \}  \;\},
\end{eqnarray}
where $x^{(i)}_r$,  and $z^{(i)}_r$ are new leaves with the same label as $c(r)$ for each $r$ in $\mbox{IR}(C)$ or $ \mbox{CR}(C)$. 
We further define:
\begin{eqnarray}
\label{ar-def}
A_r = \{x^{(i)}_r  \;|\; 1\leq i\leq |p(r)| \}.
\end{eqnarray}
for $r\in \mbox{IR}(C)$.

We now describe how to compute $f_{ux}$ defined in Eqn. (\ref{fux_def})  when $u$ is a non-leaf node in $N$.  (When $u$ is a leaf, we can computer $f_{ux}$ in the same way as in the case when $N$ is binary.)

When  $x$ is a leaf in $T$, $f_{ux}=0$ if $f_{vx}=0$ for each $v\in c(u)$.  Conversely, if $f_{v'x}=1$ for some
$f_{ux}=1$  only if  \vspace*{-0.5em}
\begin{itemize}
\item[(i)]  ${\cal L}({\cal D}_{T_C}(v)) \subseteq A(T_{C}) \cup O(T_C)$ for each $v\neq v'$ in $c(u)$, and \vspace*{-0.5em}
 \item[(ii)] $A_r\not\subseteq M_{v'x} \cup \left[ \cup_{v\in c(u)\backslash \{v'\}}{\cal L}({\cal D}_{T_C}(v)) \right]$  for each $r\in \mbox{IR}(C)$, where $M_{v'x}$ is defined in Eqn.~(\ref{mux_def}).\vspace*{-0.5em}
\end{itemize}
 The reason for (i) is that if the display of $\cD_{T}(x)$ is extended from the subtree rooted at $v'$ to 
the subtree rooted at $u$, we have to delete the subtrees rooted at any other child of $u$. The reason for (ii)
is that we cannot delete all the leaves added for each  $r \in \mbox{IR}(C)$.

It is possible that $f_{vx}=1$ for different children $v$  of $u$. However, we would like to point out,   
whether the conditions (i) and (ii)  hold or not is independent of which child is selected  when it happens.  


When $x$ is not a leaf, we assume $c(x)=\{y, z\}$. 
Clearly, if $f_{vx}=0$ for each $v\in c(u)$ and no $v'$ and $v''$ exist in $c(u)$ such that 
$f_{v'y}=1$  and $f_{v''z}=1$, then $f_{ux}=0$. 

 If $f_{v'x}=1$ for some $v'\in c(u)$, we can determine whether or not
$f_{ux}=1$ in the same way as in the case when $x$ is a leaf. 

If $f_{v'y}=1$ and $f_{v''z}=1$ for $v',  v''\in c(u)$ such that $v'\neq v''$,  
$f_{ux}=1$  only if \vspace*{-0.5em}
  \begin{itemize}
\item[(i)]  ${\cal L}({\cal D}_{T_C}(v)) \subseteq A(T_{C}) \cup O(T_C)$ for each $v\in c(u)$ such that 
$v'\neq v\neq v''$, and \vspace*{-0.5em}
 \item[(ii)] $A_r\not\subseteq M_{v'x} \cup M_{v''x} \cup  \left[ \cup_{v\in c(u) \backslash \{v',  v''\}}{\cal L}({\cal D}_{T_C}(v)) \right]$  for each $r\in \mbox{IR}(C)$. 
\end{itemize}
Again,  the two conditions are independent of which $v'$ and $v''$ are selected.

To efficiently check  the conditions (i) and (ii), we introduce some integer variables for each node.
For each $r\in \mbox{IR}(C)$ and each node $u$ in $T_C$, $m_{ur}$ denotes the number of leaves in ${\cal L}(\cD_{T_C}(u))$ that are in 
$A_r$ ; $m_{u0}$ denotes the number of non-ambiguous and non-optional leaves in ${\cal L}(\cD_{T_C}(u))$. 
It is not hard to see that $m_{u0}$ can be recursively computed using 
\begin{eqnarray}
\label{13eqn}
   m_{u0}=\left\{ \begin{array}{ll}
           1 & \mbox{if $u$ is a leaf in neither $A(T_C)$ nor $O(T_C)$},\\
           0 & \mbox{if $u$ is a leaf in $A(T_C)$ or $O(T_C)$},\\
           \sum_{v\in c(u)}m_{v0} & \mbox{if $u$ is a non-leaf node.} 
   \end{array}
\right. 
\end{eqnarray}
Similary, $m_{ur}$ can be updated as follows:
\begin{eqnarray}
\label{14eqn}
   m_{ur}=\left\{ \begin{array}{ll}
           1 & \mbox{if $u$ is a leaf in $A_{r}$},\\
           0 & \mbox{if $u$ is a  leaf not in $A_r$},\\
           \sum_{v\in c(u)}m_{vr} & \mbox{if $u$ is a non-leaf node.}
   \end{array}
\right. 
\end{eqnarray}

The subset relation in the condition (i) is equivalent to that  $m_{v0}=0$. 
Note that \\ ${\cal L}(\cD_{T_C}(u)) \cap A_r =[M_{v'x}\cup M_{v''x}\cup (\cup_{v\in c(u) \backslash \{v', v''\}}{\cal L}({\cal D}_{T_C}(v)))] \cap A_r$ for each $r\in \mbox{IR}(C)$ such that 
its leaf child $c(r)$ is not in  ${\cal L}(\cD_{T}(x))$. 
As the condition (ii) is clearly true if  
$c(r)$ is in $ {\cal L}(\cD_{T}(x))$, 
it is equivalent to 
that $m_{ur}<|A_r|$ for each $r\in \mbox{IR}(C)$ such that   $c(r)$ is not in $ {\cal L}(\cD_{T}(x))$.
Thus,  we can update  $f_{ux}$ using the formulas in Table~\ref{table2}.

\begin{table}[!t]
\begin{center}
{\small 
\begin{tabular}{c|l}
\toprule
  $x\in {\cal L}(T)$? & $f_{ux}$ and $M_{ux}$\\
\midrule
  Yes  & Case 1:   $f_{vx}=0$ for each $v\in c(u)$\\
   &~~$f_{ux}=0$;\\
  & \\
& Case 2:  $f_{v'x}=1$ for some $v'\in c(u)$\\
&  ~~If  (i) $m_{v0} =0$ for each  $v\neq v'$ in $c(u)$, and\\
&  ~~~(ii) $m_{ur}<|A_r|$ for each $r\in \mbox{IR}(C)$ 
    such that $A_{r}\cap {\cal L}(\cD_{T}(x))=\emptyset$, $\{$ \\
& ~~~~~~~~$f_{ux}=1$;\\
& ~~$\}$ else   $f_{ux}=0$;\\
\midrule
No.   & \\ 
 $c(x)=\{y, z\}$ & Case 1:   $f_{vx}=0$ for any $v\in c(u)$, and \\
& ~~~~~~~~~no $v'$ and $v''$ in $c(u)$ exist such that
  $f_{v'y}=1$ and $f_{v''z}=1$;\\
&~~$f_{ux}=0$;\\
& \\
    &   Case 2:  $f_{v'x}=1$ for some $v'\in c(u)$\\
    &   ~~ Use the same updata rule as in the case 2 when $x$ is a leaf\\
&\\
   &   Case 3:  $f_{v'y}=1$ and $f_{v''z}=1$ for some  $v'\neq v''$ in $c(u)$\\
&  ~~If (i)  $m_{v0} =0$ for each  $v\in c(u)$ such that $v'\neq v\neq v''$, and\\
&  ~~~ (ii) $m_{ur}<|A_r|$ for each $r\in \mbox{IR}(C)$ 
    such that $A_{r}\cap {\cal L}(\cD_{T}(x))=\emptyset$, $\{$ \\
& ~~~~~~~~$f_{ux}=1$;\\
& ~~$\}$ else   $f_{ux}=0$;\\
\bottomrule
\end{tabular} 
} 
\end{center}
\caption{The update rules for computing  $f_{ux}$ for a node $u$ in an arbitrary network. \label{table2}}
\end{table}

Using Eqn.~(\ref{13eqn}) and (\ref{14eqn}), we can pre-compute $m_{u0}$ and $m_{ur}$ for all $r\in \mbox{IR}(C)$ in $|{\cal E}(T_C)| \cdot (|\mbox{IR}(C)|+1)$ time.

When $f_{ux}$ is updated, we need to check whether or not $f_{vx}=1$ and $f_{vy}=1$ for each $y\in c(x)$ and
each child $v\in c(u)$. This can be done in $O(|c(u)|)$ time.  Similarly, the condition (i) is independent of $x$ and can be checked in $O(|c(u)|)$ time; the condition (ii) can be checked in $O(|\mbox{IR}(C)|)$ time.
Hence, for all nodes $x$ in $T$, the run time for updating $f_{ux}$ in $T_C$ takes 
 $O(\sum_{u\in {\cal V}(T_C)} (|c(u)| + |\mbox{IR}(C)|))=O(|{\cal E}(T_C)|+|\mbox{IR}(C)| \cdot
|{\cal V}(T_C)|)$ time.  This implies that the  run time on $T_C$ for all nodes in $T$ is 
 $O(|{\cal V}(T)|\cdot [|{\cal E}(T_C)|+ |\mbox{IR}(C)| \cdot |{\cal V}(T_C)|]\; )$ time.

Therefore, 
the total run time for determining whether or not $T$ is displayed in $N$ is 
$ O(|{\cal V}(T)|\cdot [|{\cal E}(N)|+ |{\cal R}(N)| \cdot |{\cal E}(N)|])=O(|{\cal V}(T)|\cdot |{\cal E}(N)|\cdot |{\cal R}(N)|)$ time.
 In general,  $|{\cal E}(N)|$  is not bounded  by any  function linear in the number of leaves in an arbitrary network. 

\section{A Linear-time Algorithm for the CCP}
\label{sec6}

As another application of the Decomposition Theorem, we shall design a  linear-time  algorithm for the CCP.
We first present a desired algorithm for binary reticulation-visible networks and then generalize it to non-binary networks.


\subsection{Algorithm for binary networks}

Given a binary reticulation-visible network $N$ and a subset $B\subseteq {\cal L}(N)$, the goal is to determine whether or not  $B$ is a cluster of some node in a tree displayed by $N$.

 Assume $N$ has $t$ big tree-node components
 $C_1, C_2, \cdots, C_t.$  
Consider a lowest big tree-node component $C$. 
We use the same notation as in Section~\ref{sec5}: $L_C$ is defined in Eqn.~(\ref{eq0});   $\rho(C)$ denotes  the root of $C$; $\mbox{IR}(C)$ and $\mbox{CR}(C)$ denote  the set of 
inner and cross reticulations below $C$, respectively. We also set 
$\bar{B}={\cal L}(N)\backslash B$.

When $L_C\cap B \neq \emptyset$ and $L_C\cap \bar{B} \neq \emptyset$,    $L_C$ contains $\ell_1$ and $\ell_2 $  such that 
 $\ell_1 \in B$, but $\ell_2 \not\in B$.  
  If $B$ is the cluster of a node $z$ in a subtree $T'$ of $N$,  $z$ is in the unique path $P$
from   $\rho(T')$ ($=\rho(N)$)  to $\ell_1$ in $T'$.  

Assume  $z$ is between $\rho(N)$ and  $\rho(C)$ in $P$, no matter which incoming edge is contained in $T'$ for  each $r\in \mbox{IR}(C)$, $\ell_2$ is below $\rho(C)$,     as $\rho(C)$ is visible on $\ell_2$.  This implies that $\ell_2$ is below $z$ and thus in $B$,  a contradiction.  Therefore, if  $B$ is a soft cluster, it must be  a soft cluster of a node in $C$.

When  $L_C \cap \bar{B}=\emptyset$ (that is, $L_C \subseteq B$),   we define 
\begin{equation}
\label{Eq_X}
X= \{ r\in \mbox{CR}(C) \;|\; c(r) \not\in B \}.
\end{equation}

Construct a subtree $T'$ of ${\cal D}_N(\rho(C))$ by deleting: \vspace{-0.5em}
\begin{itemize}
   \item  all but one of the incoming edges for each $r\in  \mbox{IR}(C)$,\vspace{-0.5em}
   \item all incoming edges but one with a tail not in $C$ for  each $r\in X$, and \vspace{-0.5em}
   \item  all incoming edges but one with a tail in $C$ for  each $r\in \mbox{CR}(C) \backslash  X$. \vspace{-0.5em}
\end{itemize}
We then define
\begin{equation}
\hat{B}=L_C \cup  \{ c(r) \;|\; r\in 
\mbox{CR}(C)\backslash X\}. 
\label{B}
\end{equation}
It is not hard to see that $\hat{B}$ is
 the cluster of the root of $T'$ such that $\hat{B} = B \cap {\cal L}({\cal D}_N(\rho(C))) \subseteq B$.  Hence, if $B=\hat{B}$, then $B$ is a cluster contained in ${\cal D}_N(\rho(C))$. If
$B\neq \hat{B}$, we reconstruct $N'$ from $N$ by:\vspace{-0.5em}
\begin{itemize}
 \item removing all the edges in $\{(u, r) \in {\cal E}(N) \;|\; r\in X, u\in {\cal V}(C)\}$,\vspace{-0.5em}
\item   removing all the edges in $\{(u, r) \in {\cal E}(N)  \;|\; r\in \mbox{IR}(C) \cup \mbox{CR}(C) \backslash X, u\not\in {\cal V}(C)\}$,  \vspace{-0.5em} \item  removing all but one of edges in $\{(u, r) \in {\cal E}(N)  \;|\; r\in \mbox{IR}(C) \cup \mbox{CR}(C) \backslash X, u\in {\cal V}(C)\}$, and\vspace{-0.5em}
   \item  replacing ${\cal D}_N(\rho(C))$ by a new leaf $\ell_C$.\vspace{-0.5em}
\end{itemize}
and set $B'=\{\ell_C\} \cup B\backslash \hat{B}$. We have the following fact.
\vspace{1em}

\noindent 
\begin{proposition}
\label{Prop_61}
If $L_C \cap \bar{B}=\emptyset$, 
$B$ is contained in $N$ if and only if $B'$ is contained in $N'$.
\end{proposition}

\begin{proofproof} 
Recall that 
$B' = (B  \cup \{\ell_C\}) \backslash \hat{B}$.
 If $B'$ is the cluster of a node $z$ in a tree $T''$ displayed in $N'$.
When $N'$ was reconstrcuted, $\ell_C$  replaced  the subtree $T'$ rooted at $\rho(C)$ whose leaves are $\hat{B}$; so if we re-expand $\ell_C$ into $T'$, the cluster of $z$ in $N$ becomes $\left( B \backslash \hat{B}\right)  \cup \hat{B} = B$, thus $B$ is contained in $N$.

Assume $B$ is  the cluster of a node $z$ in a subtree $T$ displayed in $N$.
Let $E$ be the set of edges entering  the reticulations nodes  such that $T=N-E$.
Since $B\neq \hat{B} = B \cap {\cal L}({\cal D}_N(\rho(C)))$, there is a leaf $\bar{\ell}\in B$  not below $\rho(C)$. Since  $\bar{\ell}$ is below $z$ in $T$,  $z$ must be above $\rho(C)$ in $T$. 

Consider a reticulation node $r\in \mbox{CR}(C)\backslash X$.  Since 
$c(r)$ is a leaf in $B$, it is  a leaf below $z$ in $T$. 
By definition of cross reticulation,  $r$ has at least one parent in $C$. Let $(p_C, r)$ be an edge such that $p_C \in C$.
Note that all but one of that incoming edges of $r$ are in $E$. 
 Define  $$E'= [ E \; \backslash \{(p_C, r) \;|\; r\in \mbox{CR}(C)\backslash X\}] \cup  \{ (p, r) \in {\cal E}(N) \;|\; r\in \mbox{CR}(C)\backslash X \; \& \; p\notin {\cal V}(C) \} $$. 
It is not hard to see that  $(p_C, r)$ is the unique incoming edge of $r$ not in $E'$ for each $r\notin X$.

 Let $T'= N - E'$.  
 It is easy to see that the cluster of $z$ is equal to $B$ and  $\hat{B}$ is the cluster of $\rho(C)$ in $T'$. Therefore, if we contract the subtree below $\rho(C)$ into a single leaf $\ell_C$,  the cluster of $z$ becomes $B \cup \{\ell_C \} \backslash \hat{B}$, which is $B'$. Therefore,  $B'$ is contained in $N'$.
\end{proofproof}

When $B\cap L_C =\emptyset$,  $B$ may or may not be  contained in ${\cal D}_N(\rho(C))$. If it is not, we use  $X$ defined in
Eqn.~(\ref{Eq_X})
to reconstruct $N'$ from $N$ by:\vspace{-0.5em}
\begin{itemize}
\item removing all the  edges in $\{(u, r)\in {\cal E}(N) \;|\; r\in \mbox{CR}(C) \mbox{ s.t. } c(r)\notin B, u\not\in  {\cal V}(C) \}$,\vspace{-0.5em}
\item   removing all the edges in $\{(u, r)\in {\cal E}(N) \;|\; r\in \mbox{CR}(C) \mbox{ s.t. } c(r)\in B,, u\in {\cal V}(C)\}$, and\vspace{-0.5em}
   \item  replacing ${\cal D}_N(\rho(C))$ by a new leaf $\ell_C$.\vspace*{-0.5em}
\end{itemize}
Similar to the last case, we have the following fact.

\begin{proposition}
\label{Prop_62}
If $B$ is not in ${\cal D}_N(\rho(C))$ and $L_C \cap B=\emptyset$,  
it is a soft cluster  in $N$  if and only if $B$ is a soft cluster in $N'$.
\end{proposition}

\begin{proofproof}
$N'$ is a subnetwork of $N$. If $B$ is a soft cluster in $N'$,  it is a soft cluster in $N$.

Conversely, assume $B$ is  the cluster of a node $z$ in a subtree $T$  of $N$.
Let  $E$ be the set of reticulation edges such that $T=N-E$.
By assumption,  $B$ is not contained in ${\cal D}_N(\rho(C))$.  Since $\rho(C)$ is visible on all leaves in $L_C$,   $\rho(C)$ is not below $z$ in $T'$.
Therefore, $\rho(C)$ and $z$ do not have ancestral relationship. 

Consider a reticulation $r \in\mbox{CR}(C)$. Since $r$ is a cross reticulation, it has a parent $p_r$ in $C$. 
If $c(r)\in B$, $c(r)$ is a leaf below $z$ in $T$ and thus  the unique  incoming edge not in $E$  has a tail not in $C$. 
If $c(r)\not\in B$,  the unique incoming edge not in $E$  may or may not have a tail in $C$.  

We  define:
$$ E'=\left(E  \cup \{(p, r) \;|\; r\in \mbox{CR}(C) \mbox{ s.t. } c(r)\not\in B, p\not\in C \}\right) \backslash\; \{(p_r, r) | r\in \mbox{CR}(C) \mbox{ s.t. } c(r)\not\in B\}.$$
Note that $(p_C, r)$ is the unique incoming edge of $r$ not in $E'$ for each $r\in  \mbox{CR}(C)$ such that  $ c(r)\not\in B$. 
Let  $T'= N - E'$. It is easy to see that the cluster of $z$ in $T'$ remains the same as the cluster of $z$ in $T$, which is equal to $B$. 
 If we contract $\mathcal{D}_{T'}(\rho(C))$ into a single leaf $\ell_C$, $T'$ is a subtree of $N'$, implying that $B$ is a soft cluster in $N'$. 
\end{proofproof}

We next show that whether or not $B$ is in $C$ can be determined in linear time.
\begin{proposition}
\label{prop63}
Let $T_C$ be a subtree constructed from $C$ in Eqn.(\ref{eq4}) and (\ref{eq5}). 
An algorithm exists that takes $T_C$ and $B$ as inputs and determines whether or not $B$ is a soft cluster in $C$ in $O(|\mathcal{E}(T_C)|)$ time.
\end{proposition}
\begin{proofproof}
First, we check whether  or not each of the leaves in $B$ is below $\rho(C)$. If  $B \not\subseteq {\cal L}(\cD_{N}(\rho(C))$,  $B$ is not a soft cluster in $D_N(\rho(C))$.  We assume that $|B|\geq 2$ and all its leaves are found below 
$\rho(C)$. 


Assume $B$ is a cluster of  a node $v$ in a subtree 
$T$ of $\cD_N(\rho(C))$.   Clearly, any network leaf in $C$ is not below $v$ if it is not in $B$.
For each $r\in \mbox{IR}(C)$, $p'\in p(r)$ exists such that $(p', r)$ was removed. If $c(r) \not\in B$, 
the parent of $r$ other than $p'$ must not be below $v$.
Since $B$ is not singleton, $v$ is an internal node in $T_C$.  Taken together, these facts imply that 
$v$ satisfies the following properties: \vspace*{-0.5em}
\begin{itemize}
\item [(i)]   Every leaf in $B$ is below $v$.\vspace*{-0.5em}
\item [(ii)]  if a leaf below $v$ is neither ambiguous nor optional, it must be in $B$.\vspace*{-0.5em}
\item [(iii)]  If  the ambiguous leaves introduced for a $r\in \mbox{IR}(C)$ are both 
  below $v$, then $c(r) \in B$.\vspace*{-0.5em}
\end{itemize} 

Conversely, if an internal node $v$ of $T_C$ satisfies the above three properties,
%
 we can then construct a subtree $T$ of $N$ in which $B$ is the cluster of $v$ as follows:\vspace{-0.5em}
\begin{itemize}
\item  For each  $r \in \mbox{IR}(C) \cup \mbox{CR}(C)$ below $v$, if it has a parent $p_1$ below $v$ and another parent $p_2$ not below $v$, remove $(p_2, r)$ if $c(r)$ is in $B$ and remove $(p_1,r)$ otherwise.
\vspace{-0.5em}
\item  For any other reticulation node not below $v$,  choose an arbitrary incoming edge to remove.
\end{itemize}
It is easy to verify that $B$ is a cluster of $v$ in the resulting tree.

Assume $B$ is a cluster of an internal node $v$ in ${\cal D}_N(\rho(C))$ and $\ell \in B$.  
If $\ell$ is not the child of an inner reticulation node below $C$, then 
$v$ is contained in the path from  $\rho(C)$ to $\ell$ in $T_C$. Otherwise, $v$ is contained in the path from  $\rho(C)$ to one of the ambiguous leaves defined for  $\ell$ in $T_C$.  

Based on the above facts, we obtain {\sc Algorithm 1} (Table~\ref{table3}) to determine whether $B$ is a cluster in $C$. 

\begin{table}[!b]
\begin{center}
\begin{tabular}{l}
 \toprule
 \hspace*{5em} {\sc Algorithm 1}\\
 Input:  $T_C$ and a subset $B$ of leaves in ${\cal D}_{N}(\rho(C))$; \smallskip\\
 
1. If $|B|=1$, {output} ``Yes" and {exit};\\
2.  Set $k=|B|$ and $m=|{\cal L}({\cal D}_{N}(\rho(C)))|-k;$\\
~~~Set a $k$-tuple $Y[1..k]$; /* {\tt record if a leaf in $B$ has been seen} */\\
~~~Set a $m$-tuple $Z[1..m]$l; \\
~~~~~ /*{\tt Use $Z$ to record the  copies of  $\ell\not\in B$ have not been seen so far} */\\
3.  Select $\ell \in B$.\\
\hspace*{1em}  {\bf for}  each leaf $x$ in $T_C$ that has the same label as  $\ell$ $\{$\\
\hspace*{2em} 3.1  $u = x$; $f=1$; /* {\tt $f$ is the no.~of leaves in $B$ have been seen so far} */\\
\hspace*{2em} 3.2  For each $i$ from $1$ to $k$, $Y[i]=0$;\\
\hspace*{2em} 3.3  For each $i$ from 1 to $m$\\
\hspace*{2em} ~~~~~if (the $i^{\mbox{\tiny th}}$ leaf $\ell' \notin B$  is ambiguous or optional) $Z[i]=2$  else $Z[i]=1$;\\
\hspace*{2em} 3.4 {\bf Repeat} while $u \neq \rho(C)$ $\{$\\
\hspace*{3.5em} $v' =u$;  $v''=\mbox{(the sibling of $u$)}$; $u= p(u)$;\\
\hspace*{3.5em} {\bf for} each $\ell'$ in the subtree \{ /* {\tt Traverse the subtree $T_C$ rooted at $v''$ }*/ \\
\hspace*{4.5em} {if} ($\ell'\in B$ having rank $j$) \& ($Y[j]== 0$) \{ $Y[j]= 1;  f = f + 1$ \};\\
\hspace*{4.5em} {if} ($\ell' \notin B$ having rank $j$) \\ 
 \hspace*{6em}   $Z[j]=Z[j]- 1$; {if} ($Z[j]==0$)   {\bf stop} Step 3.4; \\
\hspace*{3.5em} \} /* {\tt end for} */\\
\hspace*{3.5em} {if} ($f== |B|$)  {output} ``Yes" and \textbf{exit};\\
\hspace*{2em} \} /* {\tt end repeat} */\\
\hspace*{1em} \} /* {\tt end the outer for} */\\
4. output ``No" and {\bf exit};\\
\bottomrule
\end{tabular}
\caption{An algorithm to decide whether a leaf subset $B$ is a soft cluster in $C$. 
\label{table3}}
\end{center}
\end{table}

The correctness of the algorithm follows from the following facts. When the algorithm stops with answer ``No"
during the traversal of the subtree branching off at $u$ in the path from $\rho(T_C)$ to the leaf $x$. 
It means that two ambiguous copies of a leaf not in $B$ have been seen in the subtree below $u$. This implies that for any descendant $w$ of $u$, not all leaves in $B$ have been seen below $w$ and for 
$u$ and each of its ancestor, some leaf not in $B$ has two ambiguous copies below it. Hence, no node 
exists in the path from the root to $x$ in $T_C$ that satisfies the properties (i)-(iii).  
When the algorithm exists at Step 4,  $B$ is clearly not a soft cluster in $C$. 

When the algorithm stops with answer ``Yes" at $u$ inside Step 3.4,  $u$ is the lowest 
 node satisfying the three properties. Hence, $B$ is a  soft cluster in $C$.


Next, we analyze the complexity of {\sc Algorithm 1}. Step 1 and Step 2 can be done in $O(|\mathcal{L}(T_C)|)$ time. Since there are at most two leaves with that same  label as $\ell$ in $T_C$,  outer for-loop will execute twice at most. At each execution of the for-loop, Steps 3.1-3.3 take $O(|\mathcal{L}(T_C)|)$ time.
 In Step 3.4, we may traverse different subtree of $T_C$ that branch off at a node in the path from the root
of $T_C$ to $x$, so the total running time for Step 3.4 is $O(|\mathcal{E}(T_C)|)$. Hence, the algorithm runs in $O(|\mathcal{E}(T_C)| + |\mathcal{L}(T_C)|) = O(|\mathcal{V}(C)|)$, as $|\mathcal{E}(T_C)|\leq 3|\mathcal{V}(C)|$ in a binary network.

\end{proofproof}

Taking all the above facts together, we are able to give a linear time algorithm for the CCP.
\vspace*{1em}

\noindent 
\begin{center}
\begin{tabular}{l}
 \hline
 \hspace*{5em} {\sc The CCP Algorithm}\\
 {\bf Input:} A binary network $N$ and a subset $B \subseteq {\cal L}(N)$; \vspace*{0.5em} \\
1.  Compute the big tree-node components sorted in a topological order: \\
~~~~~~~~$C_1 \prec  C_2 \prec \cdots \prec  C_t$;\\
2.  {\bf for} $k=1$ {\bf to}  $t$ {\bf do} $\{$\\
\hspace*{1em} 2.1.~ Set $C=C_k$;  compute $L:=L_{C_k}$ defined in Eqn. (1);\\
 \hspace*{1em} 2.2.~ $Y := \mbox{ (~Is $B$ contained in ${\cal D}_{N}(\rho(C))$?~)}$;\\
 \hspace*{1em} 2.3.~ {\bf if} ($Y==1$) {output} ``Yes" and {\bf exit};\\
\hspace*{1em} 2.4.~ {\bf if} ($Y==0$) $\{$\\
\hspace*{3.5em}~~~ $\bar{B}:= {\cal L}(N)\backslash B$;\\
\hspace*{3.5em}~~~ {\bf if} ($L\cap \bar{B} \neq \emptyset$ \& $B\cap L \neq \emptyset$) {output} ``No" and {\bf exit};\\
\hspace*{3.5em}~~~ {\bf if} ($B\cap L==\emptyset$) $\{$\\
\hspace*{4.5em}~~~   Remove edges in $\{(u, r) \;|\; r\in \mbox{CR}(C) \mbox{ s.t. } c(r)\not\in B, u\not\in C\}$;\\
\hspace*{4.5em}~~~  Remove edges in $\{(u, r) \;|\; r\in \mbox{CR}(C) \mbox{ s.t. } c(r)\in B, u\in C\}$;\\
\hspace*{3.5em}~~~ $\}$\\
 \hspace*{3.5em}~~~ {\bf if} ($ \bar{B} \cap L==\emptyset$) $\{$\\
\hspace*{4.5em}~~~   Remove edges in $\{(u, r) \;|\; r\in \mbox{CR}(C) \mbox{ s.t. } c(r)\not\in B, u\in C\}$;\\
\hspace*{4.5em}~~~   Remove edges in $\{(u, r) \;|\; r\in \mbox{CR}(C) \mbox{ s.t. } c(r)\in B, u\not\in C\}$;\\
\hspace*{4.5em}~~~  $B:=  \left(B \cup \{\ell_{C} \}\right) \backslash \left(L \cup \{ c(r) \;|\; r\in \mbox{CR}(C) \mbox{ s.t. } c(r)\in B\}\right)$;\\
\hspace*{3.5em}~~~ $\}$\\
\hspace*{3.5em}~~~  Replace ${\cal D}_N(\rho(C))$ by a leaf $\ell_{C}$;\\
\hspace*{3.5em}~~~ Remove $C$ from the list of big tree-node components;\\
\hspace*{3.5em}~~~ Update $\mbox{CR}(C')$  for affected big tree-node components $C'$; \\ 
\hspace* {1em} ~~~$\}$\\
$\}$ /* for */\\
\hline \\
\end{tabular}
\end{center}

The obtained CCP algorithm runs in linear time.   Step 1 takes $O(|{\cal V}(N)|)=O(|{\cal L}(N)|)$ time, as $N$ is binary. 
Step 2 is a for-loop that runs $t$ times. 
Since the total number of the network leaves in $C_k$ and the reticulation nodes below $C_k$ is at most $3|{\cal V}(C_k)|$, 
Step 2.1 takes $O(|{\cal V}(C_k)|)$ time for each execution. 
In Step 2.2, the linear-time {\sc Algorithm 1}  is called to  compute  $Y$  in  
$O(|{\cal V}(C_k)|)$ time. 
Obviously, Step 2.3 takes constant time.
To implement Step 2.4  in linear time, we need to use an array $A$ to indicate whether a network leaf is in $B$ or not.
$A$ can be constructed in $O(|{\cal L}(N)|)$ time. With $A$, 
each conditional clause in Step 2.4 can be determined in $|L|$ time, which is at most 
$O(|{\cal V}(C_k)|)$. 
Since the total number of inner and cross reticulations is at most $2|{\cal V}(C_k)|$, each line of Step 2.4 takes at most $O(|{\cal V}(C_k)|)$ time.   Hence,  Step 2.4 still takes $O(|{\cal V}(C_k)|)$ time. 
Taking all these together, we have that the total time taken by Step 2 is 
$\sum_{1\leq k\leq t} O(|{\cal V}(C_k)|)
= O(|{\cal V}(N)|)=O(|{\cal L}(N)|).$
Therefore, Theorem~\ref{main-theorem2} is proved. 

\subsection{Generalization of the CCP algorithm to non-binary networks}


 Propositions 6.1 and 6.2 have been proved for non-binary reticulation-visible networks. The straightforward generalization of  {\sc Algorithm 1} does not give
a linear-time algorithm for determining whether a subset $B$ of leaves is a soft cluster in $\cD_{N}(\rho(C))$ for non-binary networks, as the outer for-loop 
in Step 3.4 will run $k$ times if $T_C$ contains $k$ ambiguous/optional leaves that have the same label as the selected leaf in $B$.  
However, it is widely known that  the lowest common ancestor (lca) of any two nodes in a tree can be computed in $O(1)$  after 
a linear-time pre-processing. In the rest of this subsection, we will use this result to prove Proposition 6.3 for non-binary networks. 

Given a non-binary  reticulation-visible network $N$ and a $B\subseteq {\cal L}(N)$ such that $|B|>1$,  We  work on a lowest big tree-node component $C$ of $N$. 
Let $T_C$ be the tree defined in Eqn.~(\ref{eq44}) and (\ref{eq55}). For each $r\in \mbox{IR}(C)$,  $A_{r}$ denotes  the set of ambiguous leaves  defined in Eqn.~(\ref{ar-def}) and $\mbox{lca}(r)$ denotes the lca of the leaves in $A_{r}$. 

\begin{proposition}
 \label{prop64}
All the nodes in $V_{\mbox{\rm lca}}=\{ \mbox{\rm lca}(r) \;|\;  r\in \mbox{IR}(C)\}$  can be computed in
 $O(|{\cal E}(N)|)$ time. 
\end{proposition}
\begin{proofproof} 
 We first pro-process $T_C$ in $O(|{\cal E}(T_C)|)$ time so that for any two nodes $u$ and $ v$ in $T_C$, $\mbox{lca}(u, v)$ can be find in $O(1)$ time \cite{Farach, Tarjan}.

Initially, each lca node is undefined. We visit  all leaves in $T_C$ in a depth-first manner.  When visiting a leaf $\ell$ that is ambiguous and added for
$r\in \mbox{IR}(C)$,  
  we set $\mbox{lca}(r)=\ell$  if  
$\mbox{lca}(r)$ is undefined,  and $\mbox{lca}(r)=\mbox{lca}(\mbox{lca}(r), \ell)$ otherwise.  Since each lca operation takes $O(1)$ time, 
the whole process takes $O(|{\cal E}(N)|)$ time. 
\end{proofproof}
 
\begin{proposition}
\label{prop65}
 {\rm  (i)} Let $\ell$ be a leaf in $T_C$ that is  neither ambiguous nor optional.  If $\ell \notin B$,  $B$ is not 
a soft cluster of any node $u$  in the path from $\rho(C)$ to $\ell$ in $N$. 

 {\rm (ii)}  For each $r\in \mbox{IR}(C)$ such that $c(r)\not\in B$,   
$B$ is not a soft cluster of any $u$  in the path from $\rho(C)$ to $\mbox{\rm lca}(r)$ inclusively in $N$. 
\end{proposition}
\begin{proofproof}
 (i) Since $\ell$ is neither ambiguous nor optional, all the nodes in the path from $\rho(C)$ to $\ell$ appears in any subtree $T$ of $N$. 
   Since $\ell\not\in B$, so, the cluster of each node in the path is not equal to $B$. 

 (ii)  For an inner reticulation node $r$ below $C$, $A_r$ contains at least two ambiguous leaves and thus $\mbox{lca}(r)$ is an internal node in $C$. Any subtree $T$ of $N$ contains exactly  one incoming edge of $r$  below $\mbox{lca}(r)$.  Thus, the cluster of each node $u$ in the path from 
$\rho(C)$ to  $\mbox{lca}(r)$ in $T$ must contain $c(r)$ and hence is not equal to $B$. 
\end{proofproof} 

Let $T_{\mbox{lca}}$ be the spanning subtree of $T_C$ over $\{ \ell \in {\cal L}(T_C) \;|\; 
\ell \not\in A(T_C)\cup O(T_C)\} \cup V_{\mbox{lca}} \cup \{\rho(C)\}$, where $A(T_C)$ and $O(T_C)$  are the set of ambiguous and optional leaves in 
$T_C$, respectively. Clearly, $T_{\mbox{lca}}$ is rooted at $\rho(C)$. 
We further define $V_{\max}=\{ v\in {\cal V}(T_C) \;|\; v\not\in {\cal V}(T_{\mbox{lca}}) \;\mbox{and}\; p(v)\in {\cal V}(T_{\mbox{lca}})\}$. 

\begin{proposition}
\label{prop66}
$B$ is a soft cluster  in $\cD_{N}(\rho(C))$ if and only if  a node $v\in V_{\max}$ exists  such that for each $\ell\in B$ there is a leaf below $v$ 
with the same label as $\ell$.
\end{proposition}
\begin{proofproof}
Assume $B$ is a soft cluster of a node $u$ in $\cD_{N}(\rho(C))$. By Proposition~\ref{prop65}, $u$ is not in $T_{\mbox{lca}}$ and thus  it is below some 
$v\in V_{\max}$. For any $\ell \in B$, $u$  and hence $v$ has a leaf descendant having the same label as $\ell$.

Let $v\in V_{\max}$ satisfy the property that for each $\ell\in B$, there is a leaf $\ell'$ having the same label as $\ell$.  
For each $x\in \mbox{IR}(C)$ such that $c(x)\not\in B$, by the definition of $V_{\max}$,  $A_x$ contains an ambiguous leaf not below $v$.
For this $x$, we select  a parent $p'_x$ of $r$ not below $v$ in $C$. 

For each $y\in \mbox{IR}(C)$ such that $c(y)  \in B$, we select a parent $p''_y$  below $v$. 

For each $r\in \mbox{CR}(C)$ such that $c(r) \in B$, we select a parent $p_r$ below $v$.

Set 
\begin{eqnarray*}
E&=&\{ (p, r)\in {\cal E}(N) \;|\; p\in {\cal V}(C), r\in A(T_C)\cup O(T_C)\}\\
          & & - \{ (p'_{x}, x) \;|\; x \in \mbox{IR}(C) \mbox{ such that }  c(x)\not\in B\}\\
      & & - \{  (p''_{y}, y) \;|\;  y\in \mbox{IR}(C) \mbox{ such that } c(r)\in B\}\\
  & & - \{  (p, r) \;|\;r\in \mbox{CR}(C) \mbox{ such that }  c(r)\in B\}
\end{eqnarray*}
Then, $\cD_{N}(\rho(C))-E$ is a subtree in which $B$ is the cluster of $v$. It is not hard to see that  $\cD_{N}(\rho(C))-E$ can be extended into a subtree of
$N$.
\end{proofproof}

Taken together, the above facts imply {\sc Algorithm 2} for determining whether a leaf subset is a soft cluster in a lowest big tree-node component or not,
presented in Table~\ref{table4}.

\begin{table}[!b]
\begin{center}
\begin{tabular}{l}
 \toprule
 \hspace*{5em} {\sc Algorithm 2}\\
 Input:  $T_C$ and a subset $B$ of leaves in ${\cal D}_{N}(\rho(C))$; \smallskip\\
 
1. If $|B|==1$, {output} ``Yes" and {\bf exit};\\
2. Construct $T_C$ defined in Eqn.~(\ref{eq44}) and (\ref{eq55});\\
3. Pre-process $T_C$ so that the lca of any two  nodes can be found in $O(1)$ time;\\
4. Traverse the leaves in $T_C$ to compute the nodes in $V_{\mbox{lca}}$; \\
5. For each  leaf $\ell \not\in A(C) \cup O(C)$ such that $\ell\not\in B$\\
~~~~~~~mark the nodes in the path from $\rho(T_C)$ to it;\\
 ~~~For each $r\in \mbox{IR}(C)$ such that $c(r) \not\in B$\\
~~~~~~~mark the nodes in the path from $\rho(T_C)$ to $\mbox{lca}(r)$ inclusively;\\
6. Traverse the nodes $u$ in $T_C$ to compute the nodes in $V_{\max}$: \\
~~~~~~  check if $u$ is unmarked and its parent is marked in Step 5 when visiting $u$;\\
7.  For each node $u\in V_{\max}$ $\{$\\
~~~~~~ 7.1  Check whether or not all leaves in $B$ are below $u$;\\
~~~~~~ 7.2  Output ``Yes" and {\bf exit} if so; \\
8. Output ``No" and {\bf exit}; \\
\bottomrule
\end{tabular}
\caption{An algorithm to decide whether $B$ is a soft cluster in $C$ in the non-binary case. }
\label{table4}
\end{center}
\end{table}

The correctness of the {\sc Algorithm 2} follows from Propositions~\ref{prop65} and \ref{prop66}. 
Step 1 takes constant time. Step 2 can be done in $O(\sum _{u\in {\cal V}(C)} |c(u)|)$ time. Step 3 takes $O(|{\cal E}(T_C)|)$ time  (see \cite{Tarjan}).
By Proposition~\ref{prop64}, Step 4 can be done in $O(|{\cal E}(T_C)|)$ time. 

Two paths from $\rho(T_C)$ to nodes in $V_{\mbox{lca}}$   may have a common subpath starting at the root. 
We mark the nodes in each of these paths in a bottom-up manner: whenever we reach a marked node, we stop the marking process in the current path.
In this way, each marked node is visited twice at most and hence Step 5 can be executed in $O(|{\cal E}(T_C)|)$ time. 

Obviously, Step 6 takes $O(|{\cal E}(T_C)|)$ time.  
For each node $u$,  Step 7.1 takes $O(|{\cal E}(\cD_{T_C}(u))|)$ time. Since  all the examined subtrees
are disjoint, the total time taken by Step 7.1 is  $O(|{\cal E}(T_C)|)$ time.  

Taken together, these facts imply that  {\sc Algorithm 2} is a line-time algorithm.
Plugging {\sc Algorithm 2} into Step 2.2 in the CCP algorithm, we can solve the CCP in linear time. 

\section{Conclusion}
\label{sec8}

Our algorithms are designed using a powerful decomposition theorem. The theorem holds for arbitrary reticulation-visible networks.  We are very interested in exploring  its applications  in the estimate of the size of a network having a visibility property and designs of algorithms for reconstructing reticulation-visible networks from gene trees or gene sequences.
Another interesting problem is  how to determine whether two networks display the same set of binary trees in polynomial time.   A solution for this is definitely valuable in phylogenetics.


\section*{Acknowledgments} 
The authors are grateful to Philippe 
Gambette, Anthony   Labarre, and St\'ephane   Vialette	for discussion on the problems studied in this work. 
This work was supported by a Singapore MOE ARF Tier-1 grant R -146-000-177-112 and the Merlion Programme 2013.


\begin{thebibliography}{10}
\providecommand{\url}[1]{\texttt{#1}}
\providecommand{\urlprefix}{URL }


\bibitem{Farach}
Bender, M. A. {\it et al.}:
 Lowest common ancestors in trees and directed acyclic graphs. {\it J. Algorithms} 57,  75--94 (2005)

\bibitem{Cardona_09_TCBB}
Cardona, G., Rossell\'o, F., Valiente, G.: Comparison of tree-child
  phylogenetic networks. {\it IEEE-ACM Trans. Comput. Biol. Bioinform.}  6,
  552--569 (2009)

\bibitem{Chan_13_PNAS}
Chan, J.M., Carlsson, G., Rabadan, R.: Topology of viral evolution. {\it PNAS}
  110,  18566--18571 (2013)




\bibitem{Dagan_08_PNAS}
Dagan, T., Artzy-Randrup, Y., Martin, W.: Modular networks and cumulative
  impact of lateral transfer in prokaryote genome evolution.{\it  PNAS}  105(29),
  10039--10044 (2008)

\bibitem{Doolittle}
Doolittle, W.F.:  Phylogenetic classification and the universal tree. {\it Science} 248, 2124--2128 (1999)
  
\bibitem{Gunawan_2015}
 Gambette, P., Gunawan, A.D.M.,   Labarre, A.,  Vialette, S.,  Zhang, L.X.:  Locating a tree in a phylogenetic network
 in quadratic time, In {\it Proc. of RECOMB}'15, pp. 96-107, Springer (2015)
 
 \bibitem{Philippe_2015}
Gambette, P., Gunawan, A.D.M.,   Labarre, A.,  Vialette, S.,  Zhang, L.X.:   Solving the tree containment problem for
genetically stable networks in quadratic time,  {\it submitted manuscript}. 

\bibitem{Johnson_Book}
Garey, M.R., Johnson, D.S.: {\it Computers and Intractability: A Guide to the Theory of NP-Completeness}. Freeman Publisher, San Francisco, USA (1979)

\bibitem{Gusfield_14_Book}
Gusfield, D.: {\it ReCombinatorics: The Algorithmics of Ancestral Recombination
  Graphs and Explicit Phylogenetic Networks}. The MIT Press (2014)

\bibitem{Tarjan}
 Harel, D.,  Tarjan, R.E.:  Fast algorithms for finding nearest common ancestors, 
{\it SIAM J. Comput.}  13: 338--355 (1984)
  
\bibitem{Huson_Recomb}
Huson, D.H., Kl\"{o}pper, T.H.: Beyond galled trees: decomposition and computation of galled networks. In: {\it Proc. of RECOMB'2007}, LNCS, vol. 4453, pp. 211-225, Springer (2007)

\bibitem{Huson_book}
Huson, D.H., Rupp, R., Scornavacca, C.: {\it Phylogenetic Networks: Concepts,
  Algorithms and Applications}. Cambridge University Press (2011)




\bibitem{Kanj_08_TCS}
Kanj, I.A., Nakhleh, L., Than, C., Xia, G.: Seeing the trees and their branches
  in the network is hard. {\it Theor. Comput. Sci.}  401,  153--164 (2008)

\bibitem{Leng_74}
 Lengauer T,  Tarjan, R.E.: A fast algorithm for finding dominators
in a flowgraph. {\it  ACM Trans.  Prog. Lang. Sys.} 1, 121--141 (1979)

\bibitem{Linz_Count}
Linz, S., John, K. S., Semple, C.: Counting trees in a phylogenetic
network is \#P-complete. {\it SIAM J. Comput.} 42, 1768--1776 (2013)

\bibitem{Marcussen_SysBiol_12}
Marcussen, T., {\it et al.}:  
, Burkhard Steuernagel, Klaus F. X. Mayer, and Odd-Arne Olsen. 
Ancient hybridizations among the ancestral genomes of bread wheat. 
{\it Science}, DOI: 10.1126/science.1250092 (2014)


\bibitem{Moret_04_TCBB}
Moret, B.M.E.,  {\it et al.}:
Phylogenetic networks: Modeling, reconstructibility, and
  accuracy. {\it IEEE-ACM TCBB.}  1,  13--23 (2004)

\bibitem{Nakhleh_13_TREE}
Nakhleh, L.: Computational approaches to species phylogeny inference and gene
  tree reconciliation. {\it Trends Ecol. Evol.}  28(12),  719--728 (2013)





\bibitem{Treangen_11_PLOSGenentics}
Treangen, T.J., Rocha, E.P.: Horizontal transfer, not duplication, drives the
  expansion of protein families in prokaryotes. {\it PLoS Genetics}  7,  e1001284
  (2011)

  
\bibitem{van_Iersel_2010_IPL}
van Iersel, L., Semple, C., Steel, M.: Locating a tree in a phylogenetic
  network. {\it Inform. Process. Letters}  110,  1037--1043 (2010)

\bibitem{Wang_01_JCB}
Wang, L., Zhang, K., Zhang, L.X.: Perfect phylogenetic networks with
  recombination. {\it J. Comp. Biol.}  8,  69--78 (2001)

\bibitem{Cui_Zhang}
Zhang, L.,  Cui, Y.:  An efficient method for DNA-based species assignment via gene tree and species tree reconciliation. In {\it Proc. of WABO'2010}, pp. 300-311. Springer (2010)

\end{thebibliography}
\end{document}